\documentclass[12pt]{article}
\usepackage{amsmath,amsfonts,latexsym,amsthm,amssymb}
\newcommand{\DM}{\mathbb D^{(\mu )}}
\newcommand{\IM}{\mathbb I^{(\mu )}}
\newcommand{\K}{\mathcal K}
\newcommand{\Rn}{\mathbb R^n}
\numberwithin{equation}{section}

\DeclareMathOperator{\I}{Im}
\DeclareMathOperator{\R}{Re}
\begin{document}
\newtheorem{teo}{Theorem}[section]
\newtheorem{lem}[teo]{Lemma}
\newtheorem{prop}[teo]{Proposition}
\newtheorem{cor}[teo]{Corollary}
\title{Distributed Order Calculus and Equations 
of Ultraslow Diffusion}
\author{Anatoly N. Kochubei
\footnote{Partially supported by the Ukrainian Foundation 
for Fundamental Research, Grant 14.1/003}
\\ \footnotesize Institute of Mathematics,\\ 
\footnotesize National Academy of Sciences of Ukraine,\\ 
\footnotesize Tereshchenkivska 3, Kiev, 01601 Ukraine}
\date{}
\maketitle
\vspace*{2cm}
Running head:\quad  ``Equations of Ultraslow Diffusion''

\newpage
\vspace*{8cm}
\begin{abstract}
We consider equations of the form
$$
\left( \mathbb D^{(\mu )}u\right) (t,x)-\Delta u(t,x)=f(t,x),
\quad t>0,x\in \mathbb R^n,
$$
where $\mathbb D^{(\mu )}$ is a distributed order derivative, that is
$$
\mathbb D^{(\mu )}\varphi (t)=\int\limits_0^1 (\mathbb D^{(\alpha )}\varphi )(t)\mu 
(\alpha )\,d\alpha ,
$$
$\mathbb D^{(\alpha )}$ is the Caputo-Dzhrbashyan fractional 
derivative of order $\alpha$, $\mu$ is a positive weight 
function.
\par
The above equation is used in physical literature for modeling 
diffusion with a logarithmic growth of the mean square 
displacement. In this work we develop a mathematical theory of 
such equations, study the derivatives and integrals of 
distributed order.

\end{abstract}

\vspace{2cm}

\newpage
\section{INTRODUCTION}

Fractional diffusion equations with the 
Caputo-Dzhrbashyan fractional time derivatives 
\begin{equation}
\left( \mathbb D^{(\alpha )}_t u\right) (t,x)-Bu(t,x)=f(t,x),\quad t>0,x\in \mathbb 
R^n,
\end{equation}
where $0<\alpha <1$, $B$ is an elliptic 
differential operator in the spatial variables,
are widely used in physics to model anomalous diffusion 
in fractal media. 

Physically, the most important characteristic of diffusion is the 
mean square displacement
$$
\overline{(\Delta x)^2}=\int\limits_{\mathbb R^n}|x-\xi 
|^2Z(t,x-\xi )\,d\xi
$$
of a diffusive particle, where $Z$ is a fundamental solution of 
the Cauchy problem for the diffusion equation. In 
normal diffusion (described by the heat equation or more general 
parabolic equations) the mean square displacement of a diffusive 
particle behaves like $\text{const}\cdot t$ for $t\to \infty$. A 
typical behavior for anomalous diffusion on some amorphous 
semiconductors, strongly porous materials etc
is $\text{const}\cdot t^\alpha$, and this was the reason to invoke the equation (1.1), 
usually with $B=\Delta$, where this anomalous behavior is an easy 
mathematical fact. There are hundreds of physical papers involving 
equations (1.1); see the surveys \cite{MK1,MK2}. The mathematical 
theory was initiated independently by Schneider and Wyss \cite{SW}
and the author \cite{K1,K2}; for more recent developments see 
\cite{EIK,EK,GM} and references therein.
\par
A number of recent publications by physicists (see 
\cite{CGS,CGSG,CKS,Nab,SCK} and references there) is devoted to the case 
where the mean square displacement 
has a logarithmic growth. This ultraslow diffusion (also called 
``a strong anomaly'') is encountered in polymer physics (a 
polyampholyte hooked around an obstacle), as well as in models of 
a particle's motion in a quenched random force field, iterated map 
models etc. In order to describe ultraslow diffusion, it is 
proposed to use evolution equations
\begin{equation}
\left( \DM_t u\right) (t,x)-Bu(t,x)=f(t,x),\quad t>0,x\in \mathbb 
R^n,
\end{equation}
where $\DM$ is the distributed order derivative (introduced 
by Caputo \cite{Cap}) of the form
\begin{equation}
\left( \DM \varphi \right) (t)=\int\limits_0^1 (\mathbb D^{(\alpha )}
\varphi )(t)\mu (\alpha )\,d\alpha ,
\end{equation}
$\mu$ is a positive weight function.

The above physical papers contain some model calculations for such 
evolution equations. There are only two mathematical papers on 
this subject. Meerschaert and Scheffler \cite{MS} developed a 
stochastic model based on random walks with a random waiting time 
between jumps. Scaling limits of these random walks are 
subordinated random processes whose density functions solve the 
ultraslow diffusion equation. The solutions in \cite{MS} are 
understood as solutions of ``algebraic'' equations obtained if the 
Laplace transform in $t$ and the Fourier transform in $x$ are 
applied.

Umarov and Gorenflo \cite{UG} applied to equations (1.2) 
Dubinskij's theory \cite{Dub} of analytic pseudo-differential 
operators. This leads to solvability results for (1.2) in the 
spaces of analytic functions and dual spaces of analytic 
functionals. Such a theory is very different from the theory of 
parabolic equations; results obtained this way ``do not feel'' the 
difference between the equations (1.2) with $B=\Delta$ and $B=-\Delta$.

The aim of this paper is to develop a theory of the model equation (1.2) 
with $B=\Delta$ comparable with the classical theory of the Cauchy 
problem for the heat equation. In particular, we construct and 
investigate in detail a fundamental solution of the Cauchy problem 
for the homogeneous equation ($f=0$) and the corresponding kernel 
appearing in the volume potential solving the inhomogeneous 
equation, prove their positivity and subordination properties. 
This leads to a rigorous understanding of a solution of the Cauchy 
problem -- it is important to know, in which sense a solution 
satisfies the equation. In its turn, this requires a deeper 
understanding of the distributed order derivative (1.3), the 
availability of its various forms resembling the classical 
fractional calculus \cite{SKM}. We also introduce and study a kind 
of a distributed order fractional integral corresponding to the 
derivative (1.3). A Marchaud-type representation of the 
distributed order derivative (based on a recent result by Samko 
and Cardoso \cite{SC}) is the main tool for obtaining, in the 
spirit of \cite{K2,EIK}, uniqueness theorems for the Cauchy 
problem for general equations (1.2) in the class of bounded 
functions and, for $n=1$ and $B=d^2/dx^2$, in the class of 
functions of sub-exponential growth. 

Comparing with the theory of 
fractional diffusion equation (1.1) we see that the distributed order
equations (under reasonable assumptions regarding $\mu$) constitute the 
limiting case equations, as $\alpha \to 0$. That is readily 
observed from estimates of fundamental solutions having, as 
$|x|\to \infty$, the estimate $\exp \left( 
-a|x|^{\frac{2}{2-\alpha }}\right)$, $a>0$, for the fractional 
diffusion equations, and $\exp (-a|x|)$ in the case of ultraslow 
diffusion.

In fact, we begin with the ``ordinary'' equation $\DM u=\lambda 
u$, $\lambda \in \mathbb R$. If $\lambda <0$, already this 
equation demonstrates a logarithmic decay of solution at infinity; 
see Theorem 2.3 below.

In general, the theory presented here is an interesting example of 
subtle analysis (with kernels from $L_1$ belonging to no $L_p$, 
$p>1$, etc) appearing in problems of a direct physical 
significance.

\section{Distributed Order Derivative}

{\bf 2.1. Definitions.} Recall that the regularized fractional 
derivative of a function $\varphi \in C[0,T]$ (also called the 
Caputo or Caputo-Dzhrbashyan derivative) of an order $\alpha \in 
(0,1)$ is defined as
\begin{equation}
\left( \mathbb D^{(\alpha )}\varphi \right) (t)=\frac{1}{\Gamma 
(1-\alpha )}\left[ \frac{d}{dt}\int\limits_0^t(t-\tau )^{-\alpha 
}\varphi (\tau )\,d\tau -t^{-\alpha }\varphi (0)\right],\quad 
0<t\le T,
\end{equation}
if the derivative in (2.1) exists. If $\varphi$ is absolutely 
continuous on $[0,T]$, then
\begin{equation}
\left( \mathbb D^{(\alpha )}\varphi \right) (t)=\frac{1}{\Gamma 
(1-\alpha )}\int\limits_0^t(t-\tau )^{-\alpha 
}\varphi' (\tau )\,d\tau
\end{equation} 
(see \cite{EIK}).

Let $\mu (t)$, $0\le t\le 1$, be a continuous non-negative 
function, different from zero on a set of a positive measure. If a 
function $\varphi$ is absolutely continuous on $[0,T]$, then by 
(1.3) and (2.2)
\begin{equation}
\left( \DM \varphi\right) (t)=\int\limits_0^tk(t-\tau )
\varphi' (\tau )\,d\tau
\end{equation} 
where
\begin{equation}
k(s)=\int\limits_0^1\frac{s^{-\alpha }}{\Gamma (1-\alpha )}\mu 
(\alpha )\,d\alpha ,\quad s>0.
\end{equation} 
It is obvious that $k$ is a positive decreasing function.

Note that for an absolutely continuous $\varphi$,
$$
\frac{d}{dt}\int\limits_0^tk(t-\tau )\varphi (\tau )\,d\tau
=\frac{d}{dt}\int\limits_0^tk(s)\varphi (t-s)\,ds
=\int\limits_0^tk(s)\varphi' (t-s)\,ds+k(t)\varphi (0),
$$
so that
\begin{equation}
\left( \DM \varphi\right) (t)=\frac{d}{dt}\int\limits_0^tk(t-\tau )
\varphi (\tau )\,d\tau -k(t)\varphi (0).
\end{equation} 

The right-hand side of (2.5) makes sense for a continuous function 
$\varphi$, for which the derivative $\dfrac{d}{dt}\int\limits_0^tk(t-\tau )
\varphi (\tau )\,d\tau$ exists. Below we use (2.5) as a general 
definition of the distributed order derivative $\DM \varphi$. 

The necessity to use the regularized fractional derivatives, not 
the Riemann-Liouville ones (defined as in (2.1), but without 
subtracting $t^{-\alpha }\varphi (0)$), in the relaxation and 
diffusion problems, is caused by the fact that a solution of an 
equation with a Riemann-Liouville derivative typically has a 
singularity at the origin $t=0$
(see, for example, \cite{EIK}), so that the initial state of a 
system to be described by the equation is not defined and requires 
a regularization. However, mathematically such problems are 
legitimate. A distributed order derivative with a constant weight, 
based on the Riemann-Liouville fractional derivative, was 
introduced by Nakhushev (see \cite{Nakh}). The diffusion equation 
with such a time derivative and a single spatial variable was 
investigated by Pskhu \cite{Pskh}. For other definitions of 
variable order and distributed order derivatives see also 
\cite{KK,LH} and references therein. In this paper we use only the 
derivatives (2.1) and (2.5).

\medskip
{\bf 2.2. Asymptotic properties.} Since the kernel (2.4) is among 
the main objects of the distributed order calculus, it is 
important to investigate its properties.

\medskip
\begin{prop}
If $\mu \in C^3[0,1]$, $\mu (1)\ne 0$, then
\begin{equation}
k(s)\sim s^{-1}(\log s)^{-2}\mu (1),\quad s\to 0,
\end{equation}
\begin{equation}
k'(s)\sim -s^{-2}(\log s)^{-2}\mu (1),\quad s\to 0.
\end{equation} 
\end{prop}

\medskip
{\it Proof}. Denote $r=-\log s$ ($\to \infty$), $\psi (\alpha )=
\dfrac{\mu (\alpha )}{\Gamma (1-\alpha )}$. Then
$$
k(s)=\int\limits_0^1\psi (\alpha )e^{r\alpha }d\alpha.
$$
Integrating twice by parts, we get
$$
k(s)=r^{-2}\int\limits_0^1\psi'' (\alpha )e^{r\alpha }d\alpha 
-r^{-1}\mu (0)-r^{-2}\left[ \psi'(1)e^r-\psi'(0)\right].
$$
We have
$$
\psi'(\alpha )=\frac{\mu'(\alpha )\Gamma (1-\alpha )+\mu (\alpha 
)\Gamma'(1-\alpha )}{[\Gamma (1-\alpha )]^2},
$$
so that $\psi'(1)=-\mu (1)$, and another integration by parts 
yields the relation
$$
k(s)=\mu (1)r^{-2}e^r+O(r^{-3}e^r),\quad r\to \infty,
$$
which implies (2.6). The proof of (2.7) is similar. $\qquad 
\blacksquare$

\medskip
It follows from (2.6) that $k\in L_1(0,T)$; however $k\notin 
L_\beta$ for any $\beta >1$. Note also that one cannot integrate 
by parts in (2.3) because, by (2.7), $k'$ has a non-integrable singularity. 

Throughout this paper we use the Laplace transform
$$
\K (p)=\int\limits_0^\infty k(s)e^{-ps}ds,\quad \R p>0.
$$
Using (2.4) and the relation
$$
\int\limits_0^\infty s^{-\alpha }e^{-ps}ds=\frac{\Gamma (1-\alpha 
)}{p^{1-\alpha }}
$$
(see 2.3.3.1 in \cite{PBM1}), we find that
\begin{equation}
\K (p)=\int\limits_0^1p^{\alpha -1}\mu (\alpha )\,d\alpha.
\end{equation}

It will often be useful to write (2.8) as
\begin{equation}
\K (p)=\int\limits_0^1e^{(\alpha -1)\log p}\mu (\alpha )\,d\alpha.
\end{equation}
Taking the principal value of the logarithm we extend $\K (p)$ to 
an analytic function on the whole complex plane cut along the 
half-axis $\mathbb R_-=\{ \I p=0,\R p\le 0\}$.

\medskip
\begin{prop}
$\mathrm{(i)}$ Let $\mu \in C^2[0,1]$. If $p\in \mathbb 
C\setminus \mathbb R_-$, $|p|\to \infty$, then
\begin{equation}
\K (p)=\frac{\mu (1)}{\log p}+O\left( (\log |p|)^{-2}\right) .
\end{equation}
More precisely, if $\mu \in C^3[0,1]$, then 
\begin{equation*}
\K (p)=\frac{\mu (1)}{\log p}-\frac{\mu'(1)}{(\log p)^2}+O\left( 
(\log |p|)^{-3}\right) .\tag{2.10$'$}
\end{equation*}

$\mathrm{(ii)}$ Let $\mu \in C[0,1]$, $\mu (0)\ne 0$. If $p\in \mathbb 
C\setminus \mathbb R_-$, $p\to 0$, then
\begin{equation}
\K (p)\sim p^{-1}\left( \log \frac{1}p\right)^{-1}\mu (0).
\end{equation}

$\mathrm{(iii)}$ Let $\mu \in C[0,1]$, $\mu (\alpha )\sim 
a\alpha^\lambda$, $a>0$, $\lambda >0$. If $p\in \mathbb 
C\setminus \mathbb R_-$, $p\to 0$, then
\begin{equation*}
\K (p)\sim a\Gamma (1+\lambda )p^{-1}\left( \log 
\frac{1}p\right)^{-1-\lambda}.\tag{2.11$'$}
\end{equation*}
\end{prop}

\medskip
{\it Proof.} (i) Integrating by parts, as in the proof of 
Proposition 2.1, we find that
$$
\int\limits_0^1e^{\alpha r}\mu (\alpha )\,d\alpha =\frac{\mu 
(1)e^r}r+O\left( r^{-2}e^r\right) ,\quad r\to \infty,
$$
which implies (2.10). The relation (2.10$'$) is proved similarly.

(ii), (iii). The relations (2.11) and (2.11$'$) follow from the 
complex version of Watson's lemma (\cite{Ol}, Chapter 4). $\qquad 
\blacksquare$

\medskip
In some cases it is convenient to use a rough estimate
$$
|\K (p)|\le C|p|^{-1}\left( \log \frac{1}{|p|}\right)^{-1},\quad |p|\le p_0,
$$
valid for any $\mu \in C[0,1]$. This estimate follows from general 
results about the behavior of the Laplace transform near the 
origin (see Chapter II, $\S 1$ of \cite{DP}).

\bigskip
{\bf 2.3. ``Ordinary'' equations.} Let us consider the simplest 
equation with a distributed order derivative, that is
\begin{equation}
\left( \DM u_\lambda \right) (t)=\lambda u_\lambda (t),\quad t>0,
\end{equation}
where $\lambda \in \mathbb R$, and it is assumed that a solution 
satisfies the initial condition $u(0)=1$. A solution of (2.12) 
should be seen as an analog of the exponential function $t\mapsto 
e^{\lambda t}$ of the classical analysis and the function 
$t\mapsto E_\alpha (\lambda t^\alpha )$, where $E_\alpha$ is the 
Mittag-Leffler function, appearing for the equation with the 
regularized fractional derivative of order $\alpha \in (0,1)$ (see 
\cite{EIK}). The equation (2.12) with $\lambda <0$ is discussed in \cite{GM1}
as the one describing distributed order relaxation. The uniqueness of a solution
will follow from the uniqueness theorem for the equation (1.2); see
Theorem 6.1. Of course the method of proof of the latter theorem can
be used to prove separately the uniqueness for a much simpler 
equation (2.12). Below we assume that $\mu \in C^2[0,1]$, $\mu 
(1)\ne 0$, $\lambda \ne 0$; evidently, $u_0(t)\equiv 1$.

Applying formally the Laplace transform to the equation (2.12) and 
taking into account the initial condition $u(0)=1$, for the 
transformed solution $\widetilde{u_\lambda }(p)$ we get
\begin{equation}
\widetilde{u_\lambda }(p)=\frac{\K (p)}{p\K (p)-\lambda }.
\end{equation}

The function (2.13) is analytic on the half-plane $\R p>\gamma$, 
if $\gamma >0$ is large enough. We have $\widetilde{u_\lambda 
}(p)\sim p^{-1}$, $p=\sigma +i\tau$, $\sigma ,\tau \in \mathbb R$, 
$|\tau |\to \infty$. Therefore \cite{DP} $\widetilde{u_\lambda }$ 
is indeed the Laplace transform of some function $u_\lambda (t)$, 
and for almost all values of $t$,
\begin{equation}
u_\lambda (t)=\frac{d}{dt}\frac{1}{2\pi i}\int\limits_{\gamma 
-i\infty}^{\gamma +i\infty }\frac{e^{pt}}{p}\frac{\K (p)}{p\K (p)-\lambda 
}\,dp.
\end{equation}

Let $\frac{1}2<\omega <1$. We will often use the contour 
$S_{\gamma ,\omega}$ in $\mathbb C$ consisting of the arc
$$
T_{\gamma ,\omega }=\{ p\in \mathbb C:\ |p|=\gamma, |\arg p|\le 
\omega \pi \},
$$
and two rays
$$
\Gamma_{\gamma ,\omega }^+=\{ p\in \mathbb C:\ |\arg p|=\omega \pi , 
|p|\ge \gamma \},
$$
$$
\Gamma_{\gamma ,\omega }^-=\{ p\in \mathbb C:\ |\arg p|=-\omega \pi , 
|p|\ge \gamma \},
$$
The  contour $S_{\gamma ,\omega}$ is oriented in the direction of 
growth of $\arg p$.

By Jordan's lemma,
$$
u_\lambda (t)=\frac{d}{dt}\frac{1}{2\pi i}\int\limits_{S_{\gamma ,\omega}}
\frac{e^{pt}}{p}\frac{\K (p)}{p\K (p)-\lambda 
}\,dp.
$$
In contrast to (2.14), here we may differentiate under the 
integral, so that
\begin{equation}
u_\lambda (t)=\frac{1}{2\pi i}\int\limits_{S_{\gamma ,\omega}}
e^{pt}\frac{\K (p)}{p\K (p)-\lambda }\,dp.
\end{equation}
Note that $\gamma$ is chosen in such a way that $p\K (p)\ne \lambda$ for all
$p\in S_{\gamma ,\omega}$ (this is possible, since $p\K (p)\to \infty$, 
as $|p|\to \infty$). 

The next result establishes qualitative properties of the function 
$u_\lambda$ resembling those of the exponential function and the 
Mittag-Leffler function. Recall that a function $u\in C^\infty 
(0,\infty )$ is called completely monotone \cite{Fe}, if 
$(-1)^nu^{(n)}(t)\ge 0$ for all $t>0$, $n=0,1,2,\ldots$.

\medskip
\begin{teo}
$\mathrm{(i)}$ The function $u_\lambda (t)$ is continuous at the 
origin $t$ and belongs to $C^\infty (0,\infty )$.

$\mathrm{(ii)}$ If $\lambda >0$, then $u_\lambda (t)$ is 
non-decreasing, and $u_\lambda (t)\ge 1$ for all $t\in (0,\infty 
)$.

$\mathrm{(iii)}$ If $\lambda <0$, then $u_\lambda (t)$ is completely 
monotone.

$\mathrm{(iv)}$ Let $\lambda <0$. If $\mu (0)\ne 0$, then
\begin{equation}
u_\lambda (t)\sim C(\log t)^{-1},\quad t\to \infty.
\end{equation}
If $\mu (\alpha )\sim a\alpha^\nu$, $\alpha \to 0$ ($a>0$, $\nu 
>0$), then
\begin{equation}
u_\lambda (t)\sim C(\log t)^{-1-\nu },\quad t\to \infty.
\end{equation}
Here and below $C$ denotes various positive constants.
\end{teo}

\medskip
{\it Proof}. The smoothness of $u_\lambda$ for $t>0$ is evident 
from (2.15). The integral in (2.15) is the sum of the integral 
over $T_{\gamma ,\omega }$ (obviously continuous at $t=0$) and the 
functions
$$
u^\pm_\lambda (t)=\frac{1}{2\pi i}\int\limits_{\Gamma^\pm_{\gamma ,\omega}}
e^{pt}\frac{\K (p)}{p\K (p)-\lambda }\,dp.
$$
We have
\begin{multline*}
u^+_\lambda (t)+u^-_\lambda (t)=\frac{1}\pi \I \left\{ e^{i\omega 
\pi }\int\limits_\gamma^\infty e^{tre^{i\omega \pi }}\frac{\K 
(re^{i\omega \pi })}{re^{i\omega \pi }\K (re^{i\omega \pi 
})-\lambda }\,dr\right\} \\
=\frac{1}\pi \I \int\limits_\gamma^\infty r^{-1}e^{tre^{i\omega \pi 
}}\,dr+\frac{\lambda }\pi \I \int\limits_\gamma^\infty \frac{e^{tre^{i\omega \pi 
}}}{r\left( re^{i\omega \pi }\K (re^{i\omega \pi 
})-\lambda \right)}\,dr.
\end{multline*}

The second summand is obviously continuous at $t=0$. The first 
summand equals
$$
\frac{1}\pi \int\limits_\gamma^\infty r^{-1}e^{tr\cos \omega \pi 
}\sin (tr\sin \omega \pi )\,dr=\frac{1}\pi \int\limits_{-\gamma t\cos 
\omega \pi}^\infty s^{-1}e^{-s}\sin (-s\tan \omega \pi )\,ds
$$
(recall that $\cos \omega \pi <0$), and this expression is continuous at $t=0$.

Let $\lambda >0$. The function $p\mapsto \dfrac{1}{p-\lambda }$ is the 
Laplace transform of the function $t\mapsto e^{\lambda t}$. Therefore \cite{Fe}
it is completely monotone. The Laplace transform of the function  
$u'_\lambda (t)$ equals
$$
p\widetilde{u_\lambda}(p)-u_\lambda (0)=\frac{p\K (p)}{p\K 
(p)-\lambda }-1=\frac{\lambda }{p\K (p)-\lambda }
$$
(strictly speaking, we have to use this formula, together with the 
asymptotics of $\K (p)$, to prove the existence of the 
Laplace transform of $u'_\lambda$).

On the other hand, the function
$$
p\K (p)=\int\limits_0^1p^\alpha \mu (\alpha )\,d\alpha
$$
is positive, while its derivative is completely monotone. By the 
Criterion 2 of the complete monotonicity (see Chapter XIII of 
\cite{Fe}), the Laplace transform of the function $u'_\lambda (t)$ 
is completely monotone. It follows from the Bernstein theorem 
about completely monotone functions and the uniqueness property 
for the Laplace transform that $u'_\lambda (t)\ge 0$ for all 
$t>0$, whence $u_\lambda$ is non-decreasing and $u_\lambda (t)\ge 
1$.

Let $\lambda <0$. Up to now, $\gamma$ was chosen so big that $p\K 
(p)\ne \lambda$ for all $p\in S_{\gamma ,\omega}$. In fact,
$$
\I p\K (p)=\int\limits_0^1|p|^\alpha \mu (\alpha )\sin (\alpha 
\arg p)\,d\alpha ,
$$
so that $\I p\K (p)=0$ only for $\arg p=0$. Meanwhile, if $\arg 
p=0$ and $\lambda <0$, then $p\K (p)-\lambda =\int\limits_0^1|p|^\alpha 
\mu (\alpha )\,d\alpha -\lambda >0$. Therefore, the above integral 
representation of $u_\lambda$ holds for any $\gamma >0$.

Since $|\K (p)|\le C|p|^{-1}\left( \log \frac{1}{|p|}\right)^{-1}$ 
for small $|p|$, we find that
$$
\left| \frac{\K (p)}{p\K (p)-\lambda }\right| \le 
C|p|^{-1}\left( \log \frac{1}{|p|}\right)^{-1}
$$
whence
$$
\left| \int\limits_{T_{\gamma ,\omega}}e^{pt}\frac{\K (p)}{p\K (p)-\lambda 
}\,dp\right| \le \frac{Ce^{\gamma t}}{\log \frac{1}\gamma }\to 0,
$$
as $\gamma \to 0$.

Considering other summands in the integral representation of 
$u_\lambda$, we see that
$$
\frac{1}\pi \int\limits_\gamma^\infty r^{-1}e^{tr e^{i\omega \pi }
}\,dr=-\frac{1}\pi \int\limits_{-\gamma t\cos 
\omega \pi}^\infty s^{-1}e^{-s}\sin (s\tan \omega \pi 
)\,ds\longrightarrow -\frac{1}\pi \int\limits_0^\infty s^{-1}e^{-s}
\sin (s\tan \omega \pi )\,ds,
$$
as $\gamma \to 0$. Next, we have to consider the expression
\begin{multline*}
\frac{\lambda }\pi \int\limits_\gamma^\infty \I \left( \frac{e^{tre^{i\omega \pi 
}}}{r}\right) \R \left( \frac{1}{re^{i\omega \pi }\K (re^{i\omega \pi 
})-\lambda }\right)\,dr\\
+\frac{\lambda }\pi \int\limits_\gamma^\infty \R \left( \frac{e^{tre^{i\omega \pi 
}}}{r}\right) \I \left( \frac{1}{re^{i\omega \pi }\K (re^{i\omega \pi 
})-\lambda }\right)\,dr\overset{\text{def}}{=}I_1+I_2.
\end{multline*}

We have
$$
\I \left( \frac{e^{tre^{i\omega \pi }}}{r}\right) =r^{-1}e^{tr\cos 
\omega \pi}\sin (tr\sin \omega \pi ),
$$
and this expression has a finite limit, as $r\to 0$. Since also 
$p\K (p)\to 0$, as $p\to 0$, we see that we may pass to the limit in 
$I_1$, as $\gamma \to 0$.

In order to consider $I_2$, we have to study the function
$$
\Phi (r,\omega )=\I \frac{1}{re^{i\omega \pi }\K (re^{i\omega \pi 
})-\lambda }.
$$
We have 
$$
re^{i\omega \pi }\K (re^{i\omega \pi })=\int\limits_0^1(re^{i\omega \pi 
})^\alpha \mu (\alpha )\,d\alpha =\int\limits_0^1 e^{-\alpha 
(s-i\omega \pi )}\mu (\alpha )\,d\alpha ,
$$
$s=-\log r\to \infty$, as $r\to 0$, so that
$$
\Phi (r,\omega )=-\frac{\int\limits_0^1 e^{-\alpha s}\sin (\alpha \omega \pi ) 
\mu (\alpha )\,d\alpha }{\left[ \int\limits_0^1 e^{-\alpha s}\cos (\alpha \omega \pi ) 
\mu (\alpha )\,d\alpha -\lambda \right]^2+\left[ \int\limits_0^1 
e^{-\alpha s}\sin (\alpha \omega \pi ) \mu (\alpha )\,d\alpha 
\right]^2}.
$$

As $s\to \infty$, the denominator tends to $\lambda^2$, while in 
the numerator
$$
\int\limits_0^1 e^{-\alpha s}\sin (\alpha \omega \pi ) 
\mu (\alpha )\,d\alpha \le C\int\limits_0^1 \alpha e^{-\alpha s} 
\le \frac{C}{s^2}=\frac{C}{(\log r)^2}.
$$
This makes it possible to pass to the limit in $I_2$, as $\gamma 
\to 0$, so that
\begin{multline}
u_\lambda (t)=-\frac{1}\pi \int\limits_0^\infty s^{-1}e^{-s}\sin 
(s\tan \omega \pi )\,ds +\frac{\lambda }\pi \int\limits_0^\infty
r^{-1}e^{tr\cos \omega \pi }\sin (tr\sin \omega \pi )\Psi 
(r,\omega )\,dr \\
+\frac{\lambda }\pi \int\limits_0^\infty
r^{-1}e^{tr\cos \omega \pi }\cos (tr\sin \omega \pi )\Phi 
(r,\omega )\,dr
\end{multline}
where
$$
\Psi (r,\omega )=\frac{\int\limits_0^1 e^{-\alpha s}\cos (\alpha \omega \pi ) 
\mu (\alpha )\,d\alpha -\lambda }{\left[ \int\limits_0^1 e^{-\alpha s}\cos (\alpha \omega \pi ) 
\mu (\alpha )\,d\alpha -\lambda \right]^2+\left[ \int\limits_0^1 
e^{-\alpha s}\sin (\alpha \omega \pi ) \mu (\alpha )\,d\alpha 
\right]^2},
$$
$s=-\log r$.

In (2.18), we may pass to the limit, as $\omega \to 1$. It is easy 
to see that the first two terms in (2.18) tend to zero, so that
\begin{equation}
u_\lambda (t)=\frac{\lambda }\pi \int\limits_0^\infty
r^{-1}e^{-tr}\Phi (r,1)\,dr,
\end{equation}
$$
\Phi (r,1)=-\frac{\int\limits_0^1 r^{\alpha }\sin (\alpha \pi ) 
\mu (\alpha )\,d\alpha }{\left[ \int\limits_0^1 r^{\alpha }\cos (\alpha \pi ) 
\mu (\alpha )\,d\alpha -\lambda \right]^2+\left[ \int\limits_0^1 
r^{\alpha }\sin (\alpha \pi ) \mu (\alpha )\,d\alpha 
\right]^2}.
$$
Since $\lambda <0$, it is seen from (2.19) that $u_\lambda$ is the 
Laplace transform of a positive function. Therefore $u_\lambda$ is 
completely monotone.

Let $\lambda <0$ and $\mu (0)\ne 0$. As we have proved, $u_\lambda$ is 
monotone decreasing. It follows from (2.13) and (2.11) that 
$\widetilde{u_\lambda }(p)\sim \dfrac{C}{p\log \frac{1}p}$, $p\to 
+0$. Applying the Karamata-Feller Tauberian theorem (see Chapter 
XIII in \cite{Fe}) we get (2.16). Similarly, if $\mu (\alpha )\sim 
a\alpha^\nu$, $\alpha \to 0$, we use the asymptotic relation 
(2.11$'$), and the same Tauberian theorem yields (2.17). $\qquad 
\blacksquare$

\medskip
A non-rigorous ``physicist-style'' proof of the statement (iii) 
was given in \cite{GM1}, where the asymptotics (2.16) was also 
found for the case $\mu (\alpha )\equiv 1$.

\section{Distributed Order Integral}

{\bf 3.1. Definition and properties.} Suppose that $\DM u=f$, 
$u(0)=0$. Applying formally the Laplace transform $u\mapsto 
\widetilde{u}$, we find that $\widetilde{u}(p)=\dfrac{1}{p\K 
(p)}\widetilde{f}(p)$. The asymptotic properties of $\K (p)$ show 
(see \cite{DP}) that the function $p\mapsto \dfrac{1}{p\K (p)}$ is the 
Laplace transform of some function $\varkappa (t)$, and
\begin{equation}
\varkappa (t)=\frac{d}{dt}\frac{1}{2\pi i}\int\limits_{\gamma 
-i\infty }^{\gamma +i\infty }\frac{e^{pt}}{p}\cdot \frac{1}{p\K (p)}\,dp,
\quad \gamma >0.
\end{equation}

It is natural to define {\it the distributed order integral} $\IM$, as the
convolution operator
$$
\left( \IM f\right) (t)=\int\limits_0^t\varkappa (t-s)f(s)\,ds.
$$

\medskip
\begin{prop}
Suppose that $\mu \in C^3[0,1]$, $\mu (1)\ne 0$, and either $\mu 
(0)\ne 0$ or $\mu (\alpha )\sim a\alpha^\nu$, $a>0$, $\nu >0$. 
Then:

$\mathrm{(i)}$ $\varkappa \in C^\infty (0,\infty )$, and 
$\varkappa$ is completely monotone;

$\mathrm{(ii)}$ for small values of $t$,
\begin{equation}
\varkappa (t)\le C\log \frac{1}{t},
\end{equation} 
\begin{equation}
|\varkappa' (t)|\le Ct^{-1}\log \frac{1}{t},
\end{equation} 
\end{prop}

\medskip
{\it Proof}. As in Sect. 2, we deform the contour of integration 
in (3.1) and differentiate:
\begin{equation}
\varkappa (t)=\frac{1}{2\pi i}\int\limits_{S_{\gamma ,\omega 
}}\frac{e^{pt}}{p\K (p)}\,dp.
\end{equation}

We will need information about the asymptotic behavior of $\dfrac{1}{\K (p)}$.
By (2.10$'$),
$$
\K (p)=\frac{\mu (1)}{\log p}-\frac{\mu' (1)}{(\log p)^2}+c(p),
\quad c(p)=O\left( \frac{1}{(\log |p|)^3}\right) ,\quad |p|\to 
\infty .
$$
Then we can write
$$
\frac{1}{\K (p)}-\frac{\log p}{\mu (1)}-\frac{\mu'(1)}{[\mu 
(1)]^2}=\frac{-\mu (1)c(p)(\log p)^3+[\mu'(1)]^2-\mu'(1)c(p)(\log 
p)^2}{[\mu (1)]^2[\mu (1)\log p-\mu'(1)+c(p)(\log p)^2]},
$$
whence
\begin{equation} 
\frac{1}{\K (p)}=\frac{\log p}{\mu (1)}+\frac{\mu'(1)}{[\mu 
(1)]^2}+O\left( \frac{1}{(\log |p|)}\right) ,\quad p\to \infty .
\end{equation}

The integral in (3.4) consists of the integral over $T_{\gamma 
,\omega}$ (a function from $C^\infty [0,\infty )$) and integrals 
over $\Gamma^\pm_{\gamma ,\omega}$. Each of the latter ones is 
estimated, due to (3.5), by an expression
$$
C\int\limits_\gamma^\infty e^{-art}r^{-1}\log r\,dr\sim C\log 
\frac{1}t,\quad t\to 0
$$
($a,C>0$; see the asymptotic formula (13.49) in \cite{Ri1}). This implies
(3.2). The proof of (3.3) is similarly based on the same asymptotic relation
from \cite{Ri1}.

In order to prove that $\varkappa$ is completely monotone, we 
proceed as in the proof of Theorem 2.3, to transform (3.4) into a 
representation by a Laplace integral. First we pass to the limit, 
as $\gamma \to 0$. This is possible because, by Proposition 2.2, 
either
\begin{equation} 
\frac{1}{p\K (p)}\sim \mu (0)\log \frac{1}p,\quad p\to 0,
\end{equation}
if $\mu (0)\ne 0$, or
\begin{equation} 
\frac{1}{p\K (p)}\sim C\left( \log \frac{1}p\right)^{1+\nu },\quad p\to 0,
\end{equation}
if $\mu (\alpha )\sim a\alpha^\nu$, $\alpha \to 0$. Both the 
relations (3.6) and (3.7) are sufficient to prove that the 
integral over $T_{\gamma ,\omega}$ tends to 0, as $\gamma \to 0$, 
while the $\gamma \to 0$ limits of both the integrals 
over $\Gamma^\pm_{\gamma ,\omega}$ exist. We come to the 
representation
\begin{equation}
\varkappa (t)=\frac{1}\pi \I \left\{ e^{i\omega \pi 
}\int\limits_0^\infty e^{tre^{i\omega \pi }}\frac{dr}{re^{i\omega \pi 
}\K (re^{i\omega \pi })}\right\}.
\end{equation}

We find, introducing the parameter $s=-\log r\to \infty$, as $r\to 
0$, that
\begin{multline*}
re^{i\omega \pi }\K (re^{i\omega \pi })=\int\limits_0^1\left( re^{i\omega \pi}
\right)^\alpha \mu (\alpha )\,d\alpha =\int\limits_0^1 e^{-\alpha 
(s-i\omega \pi )}\mu (\alpha )\,d\alpha \\
=\int\limits_0^1 e^{-\alpha s}(\cos (\alpha \omega \pi )+i\sin
(\alpha \omega \pi ))\mu (\alpha )\,d\alpha .
\end{multline*}
Taking into account the logarithmic behavior of the integrand of 
(3.8) near the origin, we may pass to the limit in (3.8), as 
$\omega \to 1$, and we get that
$$
\varkappa (t)=\frac{1}\pi \int\limits_0^\infty e^{-tr}
\frac{\int\limits_0^1 r^\alpha \sin (\alpha \pi )\mu (\alpha 
)\,d\alpha }{\left[ \int\limits_0^1 r^\alpha \cos (\alpha \pi )\mu (\alpha 
)\,d\alpha \right]^2+\left[ \int\limits_0^1 r^\alpha \sin (\alpha \pi )\mu (\alpha 
)\,d\alpha \right]^2}\,dr,
$$
as desired. $\qquad \blacksquare$

\medskip
Note that, by (3.2), $\varkappa \in L_1^{\text{loc}}(0,\infty )$.

\medskip
{\bf 3.2. The Marchaud form of the distributed order derivative.} 
If $f\in L_1(0,T)$, $u=\IM f$, then $u=\varkappa *f$,
$$
\left( \DM u\right) (t)=\frac{d}{dt}(k*\varkappa 
*f)(t)=\frac{d}{dt}(1*f)=\frac{d}{dt}\int\limits_0^tf(\tau 
)\,d\tau =f(t)
$$
almost everywhere. Thus $\DM \IM =I$ on $L_1(0,T)$.

The identity $(k*\varkappa )(t)\equiv 1$ (almost everywhere), 
which follows from the fact that the product of the Laplace 
transforms $\K (p)$ and $\frac{1}{p\K (p)}$ equals $\frac{1}{p}$, 
means that $\varkappa$ is {\it a Sonine kernel} (see \cite{SC}). 
Since both the functions $k$ and $\varkappa$ are monotone 
decreasing (obviously, $k$ is completely monotone), we are within 
the conditions \cite{SC}, under which the operator $\DM$, on 
functions $u=\IM f$, $f\in L_p(0,T)$, $1<p<\infty$, can be 
represented in the form
\begin{equation}
\left( \DM u\right) (t)=k(t)u(t)+\int\limits_0^tk'(\tau )[u(t-\tau 
)-u(t)]\,d\tau ,\quad 0<t\le T,
\end{equation}
where the representation (3.9) is understood as follows. Let
$$
\left( \Psi_\varepsilon u\right) (t)=\begin{cases}
\int\limits_\varepsilon^tk'(\tau )[u(t-\tau 
)-u(t)]\,d\tau, & \text{if $t\ge \varepsilon$},\\
0, & \text{if $0<t<\varepsilon$}.\end{cases}
$$
Then
\begin{equation}
\lim\limits_{\varepsilon \to 0}\left\| \left( \DM u\right) (t)
-k(t)u(t) -\left( \Psi_\varepsilon u\right) 
(t)\right\|_{L_p(0,T)}=0.
\end{equation}

The representation (3.9), similar to the Marchaud form of a 
fractional derivative \cite{SKM}, will be useful for our proofs of 
uniqueness theorems, because the integral operator in (3.9) has 
the form enabling the maximum principle approach. On the other 
hand, the precaution we made understanding (3.9) in terms of 
(3.10) cannot be easily avoided due to a strong singularity of 
$k'$ in accordance with the asymptotics (2.7).
 
\section{The Equation of Ultraslow Diffusion}

{\bf 4.1. A fundamental solution of the Cauchy problem.} Let us 
consider the equation (1.2) with $B=\Delta$, that is
\begin{equation}
\left( \DM_t u\right) (t,x)=\Delta u(t,x),\quad x\in \Rn ,t>0.
\end{equation}
In this section we construct fundamental solution $Z(t,x)$ of the Cauchy 
problem, a solution of (4.1) with $Z(0,x)=\delta (x)$, and obtain 
its estimates.

Below we use the following normalization of the Fourier transform:
$$
\widehat{u}(\xi )=\int\limits_{\Rn}e^{ix\cdot \xi}u(x)\,dx,
$$
so that
$$
u(x)=\frac{1}{(2\pi )^n}\int\limits_{\Rn}e^{-ix\cdot \xi}\widehat{u}(\xi 
)\,d\xi .
$$
For a radial function $u(r)$, $r=|x|$,
\begin{equation}
\widehat{u}(r)=2\pi^{n/2}\left( \frac{r}2\right)^{1-\frac{n}2}
\int\limits_0^\infty \rho^{n/2}u(\rho )J_{\frac{n}2-1}(r\rho 
)\,d\rho ,
\end{equation}
where $J_\nu$ is the Bessel function.

Applying formally the Laplace transform in $t$ and the Fourier 
transform in $x$, we find that
$$
\widehat{\widetilde{Z}} (p,\xi )=\frac{\K (p)}{p\K (p)+|\xi |^2}.
$$
By (4.2),
\begin{equation}
\widetilde{Z}(p,x)=(2\pi )^{-\frac{n}2}|x|^{1-\frac{n}2}\K (p)
\int\limits_0^\infty \frac{s^{n/2}}{p\K 
(p)+s^2}J_{\frac{n}2-1}(|x|s)\,ds.
\end{equation}

It is known (\cite{PBM2}, 2.12.4.28) that
\begin{equation}
\int\limits_0^\infty \frac{y^{\nu +1}}{y^2+z^2}J_\nu (cy)\,dy=z^\nu 
K_\nu (cz),\quad -1<\nu <\frac{3}2,
\end{equation}
where $K_\nu$ is the McDonald function. If $n\le 4$, then the 
above restriction upon $\nu =\dfrac{n}2-1$ is satisfied, and (4.4) 
implies the representation
\begin{equation}
\widetilde{Z}(p,x)=(2\pi )^{-\frac{n}2}|x|^{1-\frac{n}2}\K (p)
(p\K (p))^{\frac{1}2(\frac{n}2-1)}K_{\frac{n}2-1}(|x|\sqrt{p\K 
(p)}).
\end{equation}
We have simpler formulas in the lowest dimensions -- if $n=2$, 
then
\begin{equation}
\widetilde{Z}(p,x)=\frac{1}{2\pi }\K (p)K_0(|x|\sqrt{p\K 
(p)});
\end{equation}
if $n=1$, then
\begin{equation}
\widetilde{Z}(p,x)=\frac{1}{2}\frac{\K (p)}{\sqrt{p\K (p)}}e^{-|x|\sqrt{p\K 
(p)}}
\end{equation}
because $K_{-1/2}(z)=K_{1/2}(z)=\sqrt{\frac{\pi }{2z}}e^{-z}$ (see 
\cite{BE}).

The function $K_\nu$ decays exponentially at infinity: 
$$
K_\nu (z)\sim \sqrt{\frac{\pi }{2z}}e^{-z},\quad z\to \infty ,
$$
while $K_\nu (z)\sim Cz^{-\nu }$, as $z\to 0$ (if $\nu >0$), and 
$K_0(z)\sim -\log z$. We see that the function on the right in 
(4.5) belongs to $L_1(\Rn )$ in $x$ for any $n$, not only for 
$n\le 4$. Using the identity
$$
\int\limits_0^\infty rJ_\nu (br)K_\nu (cr)\,dr=b^\nu c^{-\nu 
}(b^2+c^2)^{-1}
$$
(\cite{PBM2}, 2.16.21.1) we check that the inverse Fourier 
transform of the right-hand side of (4.5) coincides with $\K 
(p)\left( p\K (p)+|\xi |^2\right)^{-1}$. Therefore the formula 
(4.5) is valid for any $n$.

Let us consider estimates of the function $Z$ and its derivatives. 
Qualitatively, the behavior of $Z$ is similar to that of the
fundamental solution of the Cauchy problem for the fractional 
diffusion equation (1.1) (see \cite{K2,EK,EIK}). In addition to 
the singularity at $t=0$, $Z(t,x)$ has, if $n>1$, a singularity at 
$x=0$ (a logarithmic singularity, if $n=2$, and a power one, if 
$n\ge 3$). As usual $Z(t,x)\to \delta (x)$, as $t\to 0$. This 
means that the singularity at $t=0$ becomes ``visible'' near the 
origin in $x$. In fact, we obtain separate estimates for a small 
$|x|$, showing the character of singularities in $t$ and $x$, and 
for a large $|x|$. In addition, subsequent applications of the 
fundamental solutions require estimates of $\DM Z$, applicable 
simultaneously for all $x\ne 0$, and uniform in $t$. Of course, 
estimates for $\DM Z$ at the origin and infinity can be obtained 
from the relation $\DM Z=\Delta Z$.

All the above estimates deal with a finite time interval, $t\in (0,T]$, 
and it is this kind of estimates, that is needed to study the Cauchy
problem. Separately we will give some estimates of $Z$ for large values 
of $t$, just to see the qualitative behavior of $Z$.

\medskip
\begin{teo}
Suppose that $\mu \in C^2[0,1]$, $\mu (\alpha )=\alpha^\nu \mu_1 (\alpha )$,
$\mu_1(\alpha )\ge \rho >0$, $0\le \alpha \le 1$, $\nu \ge 0$. 
Denote by $\varepsilon$ a small positive number. The function $Z$ is infinitely
differentiable for $t\ne 0$ and $x\ne 0$. The following estimates hold
for $0<t\le T$.

If $n=1$, then
\begin{equation}
\left| D_x^mZ(t,x)\right| \le Ct^{-\frac{m+1}2},\quad |x|\le 
\varepsilon ,0\le m\le 3. 
\end{equation}

If $n=2$, then
\begin{equation}
\left| Z(t,x)\right| \le Ct^{-1}\log |x|^{-1},\quad |x|\le\varepsilon , 
\end{equation}
\begin{equation}
\left| D_x^mZ(t,x)\right| \le Ct^{-1}|x|^{-m},\quad |x|\le 
\varepsilon ,1\le m\le 3. 
\end{equation}

If $n\ge 3$, then
\begin{equation}
\left| D_x^mZ(t,x)\right| \le Ct^{-1}|x|^{-n+2-m},\quad |x|\le 
\varepsilon ,0\le m\le 3. 
\end{equation}

In all cases,
\begin{equation}
\left| D_x^mZ(t,x)\right| \le Ce^{-a|x|}\ (a>0),\quad |x|\ge 
\varepsilon^{-1}.
\end{equation}

The estimate of $\DM Z$, uniform in $t$, is as follows:
\begin{equation}
\left| \left( \DM Z\right) (t,x)\right| \le C|x|^{-n-2}e^{-a|x|}\ 
(a>0),\quad |x|\ne 0.
\end{equation}

If $|x|\le \varepsilon$, then
\begin{equation*}
\left| \left( \DM Z\right) (t,x)\right| \le Ct^{-2}|x|^{-n+2}. 
\tag{4.13$'$}
\end{equation*}
\end{teo}

\bigskip
{\it Proof}. As before, using Jordan's lemma we write
\begin{equation}
Z(t,x)=(2\pi )^{-\frac{n}2}|x|^{1-\frac{n}2}
\int\limits_{S_{\gamma ,\omega }}e^{pt}\K (p)
(p\K (p))^{\frac{1}2(\frac{n}2-1)}K_{\frac{n}2-1}(|x|\sqrt{p\K 
(p)})\,dp,\quad x\ne 0.
\end{equation}
The integral in (4.14) consists of the ones on $T_{\gamma ,\omega 
}$ and $\Gamma^\pm_{\gamma ,\omega }$. Let us begin with the first 
of them, denoted by $Z^0(t,x)$. Below we assume that $\gamma >e$.

If $p\in T_{\gamma ,\omega }$, then $p=\gamma e^{i\varphi }$, 
$|\varphi |\le \omega \pi$, $\frac{1}2\le \omega <1$. Under our 
assumptions,
$$
p\K (p)=\int\limits_0^1\gamma^\alpha e^{i\alpha \varphi 
}\alpha^\nu \mu_1(\alpha )\,d\alpha .
$$
Let us consider the location of values of $p\K (p)$, $p\in T_{\gamma ,\omega 
}$. If $|\varphi |\le \pi /2$, then $\R p\K (p)\ge 0$. Suppose 
that $\frac{\pi }2<\varphi \le \omega \pi$. Then $\R p\K (p)\ge 
R\cos (\omega \pi )$, $R=\int\limits_0^1\gamma^\alpha \alpha^\nu \mu_1(\alpha )\,d\alpha 
$,
$$
\I p\K (p)\ge \rho \int\limits_0^1\alpha^\nu \sin (\alpha \varphi )\,d\alpha 
=\rho \varphi^{-1-\nu }\int\limits_0^\varphi \beta^\nu \sin \beta 
\,d\beta \ge \rho (\omega \pi )^{-1-\nu }\int\limits_0^{\pi /2}\beta^\nu \sin \beta 
\,d\beta >0,
$$
so that $0\le \arg p\K (p)<\pi$, as $p$ belongs to the part of $T_{\gamma ,\omega }$
lying in the upper half-plane. Similarly, $-\pi <\arg p\K (p)\le 
0$ for the part from the lower half-plane. Thus, $|\arg p\K 
(p)|<\pi$, and since $T_{\gamma ,\omega }$ is compact, we have $|\arg p\K 
(p)|\le \varphi_0<\pi$, $p\in T_{\gamma ,\omega }$.

This means that
$$
\R \sqrt{p\K (p)}\ge \cos \frac{\varphi_0}2\cdot 
\inf\limits_{p\in T_{\gamma ,\omega }}\left| \int\limits_0^1p^\alpha 
\mu (\alpha )\,d\alpha \right| \overset{\text{def}}{=}r_0>0
$$
because $\I  \int\limits_0^1p^\alpha \mu (\alpha )\,d\alpha =0$ 
with $p\in T_{\gamma ,\omega }$ only if $p=\gamma$, and there $\R
\int\limits_0^1\gamma^\alpha \mu (\alpha )\,d\alpha \ne 0$.

Therefore, using the above-mentioned asymptotics of the McDonald 
function, we find that
\begin{equation}
\left| Z^0(t,x)\right| \le Ce^{-a|x|}\ (a>0),\quad |x|\ge 
\varepsilon^{-1}.
\end{equation}
As $|x|\le \varepsilon$, we get
\begin{equation}
\left| Z^0(t,x)\right| \le \begin{cases}
C, & \text{if $n=1$;}\\
C\log |x|^{-1}, & \text{if $n=2$;}\\ 
C|x|^{-n+2}, & \text{if $n\ge 3$.}\end{cases}
\end{equation}

Let $Z^\pm (t,x)$ be the parts of $Z(t,x)$ corresponding to the 
integration over $\Gamma_{\gamma ,\omega}^\pm$. If, for example, 
$n\ge 3$, then
\begin{equation}
\left| Z^\pm (t,x)\right| \le 
C|x|^{\frac{1-n}2}\int\limits_\gamma^\infty e^{tr\cos (\omega \pi 
)}\frac{1}{\log r}\left( \frac{r}{\log r}\right)^{\frac{n-3}4}
e^{-a|x|\left( \frac{r}{\log r}\right)^{1/2}}dr,\quad a>0.
\end{equation}

Let us make the change of variables $z=z(r)=\left( \dfrac{r}{\log 
r}\right)^{1/2}$. In order to express (asymptotically) $r$ as a 
function of $z$, we denote $s=\log r$, so that $s^{-1}e^s=z^2$ 
where $s\to \infty$ and $z\to \infty$. Taking the logarithm of 
both parts of the last equality we get $s-\log s=2\log z$. It is 
known (\cite{Fed}, page 50) that
$$
s=2\log z+O(\log \log z),\quad z\to \infty .
$$
Therefore $r=r(z)$ satisfies the inequalities
\begin{equation}
z^2(\log z)^{-b}\le r(z)\le (\log z)^b
\end{equation}
for some $b\ge 0$.

For $|x|\ge \varepsilon^{-1}$, the factor $e^{tr\cos (\omega \pi 
)}$ in (4.17) can be estimated by 1, and after the use of (4.18) 
the power terms, as well as the logarithmic ones, are dominated by 
the exponential factor (the integral in $z$ is taken over 
$(\gamma_1,\infty )$, $\gamma_1>0$), so that 
$$
\left| Z^\pm (t,x)\right| \le Ce^{-a'|x|},\quad a'>0,
$$
and, together with (4.15), this implies (4.12) for $n\ge 3$, 
$m=0$.

For $|x|<\varepsilon$, the factor $e^{-a|x|\left( \frac{r}{\log 
r}\right)^{1/2}}$ is estimated by 1, and an elementary estimate 
gives that
$$
\left| Z^\pm (t,x)\right| \le Ct^{-1}|x|^{-n+2},
$$
which implies the required estimate of $Z$.

The bounds for the derivatives, as well as the estimates 
(4.8)-(4.10) for $n=1$ and $n=2$ are obtained in a similar way. 
Some of the estimates can in fact be slightly refined (using the 
asymptotic formulas for the Laplace integrals with logarithmic 
factors \cite{Ri1}), involving $\dfrac{t^{-1}}{\log t^{-1}}$ for 
small values of $t$, instead of $t^{-1}$.

Let us prove (4.13). Let $n\ge 3$; the cases $n=2$ and $n=1$ are 
similar. The estimates of the McDonald function for small and 
large arguments can be combined as follows:
\begin{equation}
\left| K_{\frac{n}2-1}(z)\right| \le 
C|z|^{-\frac{n}2+1}e^{-a|z|},\quad z\ne 0,
\end{equation}
with a possibly different choice of the constant $a>0$.

Next, let us write an integral representation of $\DM Z$. If 
$u(t)=e^{pt}$, then
$$
\left( \DM u\right) (t)=p\int\limits_0^tk(t-\tau )e^{p\tau 
}\,d\tau =pe^{pt}\int\limits_0^tk(s)e^{-ps}\,ds.
$$
Using the expression (2.4) for $k(s)$, the identity 
$$
\int\limits_0^ts^{-\alpha }e^{-ps}\,ds=p^{\alpha -1}\gamma 
(1-\alpha ,pt)
$$
(\cite{PBM1}, 1.3.2.3), where $\gamma$ is an incomplete gamma 
function, we find that
$$
\left( \DM u\right) (t)=e^{pt}\int\limits_0^1p^\alpha \frac{\mu 
(\alpha)}{\Gamma (1-\alpha )}\gamma (1-\alpha ,pt)\,d\alpha .
$$
It is known that
$$
\gamma (1-\alpha ,z)\sim \frac{1}{1-\alpha }z^{1-\alpha },\quad 
z\to 0,
$$
$$
\gamma (1-\alpha ,z)\sim \Gamma (1-\alpha 
)-z^{-\alpha}e^{-z},\quad z\to \infty
$$
(see Chapter 9 in \cite{BE}). This implies the inequality
$$
\left| \frac{\gamma (1-\alpha ,z)}{\Gamma (1-\alpha )}\right| \le 
C
$$
valid, in particular, for all $z=pt$, $p\in S_{\gamma ,\omega }$, 
$t\in (0,T]$. Recalling also that
$$
\int\limits_0^1p^\alpha \mu (\alpha )\,d\alpha =p\K (p)
$$
we see that the application of $\DM$ to the integral representing 
$Z$ leads to the appearance of the factor $|p\K (p)|$ in the 
estimates of $\DM Z$, compared to those of $Z$. Using also (4.19) 
and estimating by 1 the decaying exponential involving $t$ (in the 
integrals over $\Gamma^\pm_{\gamma ,\omega }$) we come to the 
inequality (4.13).

The proof of the inequality (4.13$'$) is similar to those for 
estimates of the derivatives in spatial variables. $\qquad 
\blacksquare$

\bigskip
{\bf 4.2. Subordination and positivity.} Let us find a connection 
between $Z$ and the fundamental solution of the heat equation. Our 
approach follows \cite{CGSG} where the case $n=1$ was considered 
(without a full rigor).

Let us consider the function
$$
g(u,p)=\K (p)e^{-up\K (p)},\quad u>0,\R p>0.
$$
Let $p=\gamma +i\tau$, $\gamma >0$, $\tau \in \mathbb R$. As 
$|\tau |\to \infty$,
$$
\K (p)\sim \frac{C}{\log \sqrt{\gamma^2+\tau^2}+i\arg p},\quad 
\arg p\to \pm \frac{\pi }2.
$$
It follows that
$$
\R (p\K (p))\sim C\left\{ \frac{\gamma }{\log \sqrt{\gamma^2+\tau^2}}+\frac{\pi 
}2\frac{|\tau |}{(\log \sqrt{\gamma^2+\tau^2})^2}\right\},\quad |\tau |\to 
\infty ,
$$
whence
\begin{equation}
\left| e^{-up\K (p)}\right| \le C\exp \left\{-au\left( \frac{\gamma }{\log (\gamma^2+\tau^2)}
+\frac{|\tau |}{(\log (\gamma^2+\tau^2))^2}\right)\right\}
,\quad a>0.
\end{equation}

Writing $\log (\gamma^2+\tau^2)\le 
C(\gamma^2+\tau^2)^\varepsilon$, $0<\varepsilon <1/4$, we find 
from (4.20) that
\begin{multline*}
\int\limits_{-\infty }^\infty |g(u,\gamma +i\tau )|\,d\tau \le
C\int\limits_0^\infty \exp \left( -au\left( \frac{\gamma }{(\gamma^2+\tau^2)^\varepsilon}
+\frac{\tau }{(\gamma^2+\tau^2)^{2\varepsilon}}\right) \right) 
\,d\tau \\
\le C\int\limits_0^1e^{-au\frac{\gamma 
}{(\gamma^2+\tau^2)^\varepsilon}}\,d\tau +C\int\limits_1^\infty
e^{-au\frac{\tau }{(\gamma^2+\tau^2)^{2\varepsilon}}}\,d\tau
\le Ce^{-au\frac{\gamma }{(\gamma^2+1)^\varepsilon }} 
+C\gamma \int\limits_{\gamma^{-1}}^\infty 
e^{-a'u\gamma^{1-4\varepsilon}y^{1-4\varepsilon}}\,dy\\
\le C+C\int\limits_0^\infty e^{-a'uz^{1-4\varepsilon}}\,dz
\end{multline*}
($a'>0$), whence
\begin{equation}
\sup\limits_{\gamma \ge 1}\int\limits_{-\infty }^\infty |g(u,\gamma +i\tau 
)|\,d\tau <\infty .
\end{equation}

It follows from (4.21) (see \cite{DP}) that $g(u,p)$ is the 
Laplace transform of some locally integrable function $G(u,t)$:
\begin{equation}
g(u,p)=\int\limits_0^\infty e^{-pt}G(u,t)\,dt,
\end{equation}
and the integral in (4.22) is absolutely convergent if $\R p\ge 
1$.

On the other hand, the function $p\K (p)$ is positive and has a 
completely monotone derivative, so that $e^{-up\K (p)}$ is 
completely monotone. Since $\K (p)$ is completely monotone, we 
find that $g$ is completely monotone in $p$ (we have used Criteria 
1 and 2 of the complete monotonicity; see \cite{Fe}), so that 
$G(u,t)\ge 0$ by Bernstein's theorem.

\medskip
\begin{teo}
$\mathrm{(i)}$ The fundamental solution $Z(t,x)$ satisfies the subordination
identity
\begin{equation}
Z(t,x)=\int\limits_0^\infty G(u,t)(4\pi u)^{-n/2}e^{-\frac{|x|^2}{4u}}\,du,
\quad x\ne 0,t>0,
\end{equation} 
where $G(u,t)\ge 0$ and
\begin{equation}
\int\limits_{\Rn} G(u,t)\,du=1.
\end{equation} 

$\mathrm{(ii)}$ For all $t>0$, $x\ne 0$, $Z(t,x)$ is non-negative, 
and
\begin{equation}
\int\limits_{\Rn} Z(t,x)\,dx=1.
\end{equation} 
\end{teo}

\medskip
{\it Proof.} In order to prove (4.24), we integrate (4.22) in $p$ 
using Fubini's theorem. We get
$$
\int\limits_0^\infty e^{-pt}\,dt\int\limits_0^\infty 
G(u,t)\,du=\frac{1}p,
$$
which implies (4.24).

Let us prove (4.23). The convergence of the integral at infinity 
follows from (4.24), while near the origin the function $u\mapsto 
(4\pi u)^{-n/2}e^{-\frac{|x|^2}{4u}}$ decays exponentially. Let 
$v(t,x)$ be the right-hand side of (4.23). Multiplying by 
$e^{-pt}$ and integrating in $t$ we find that
$$
\int\limits_0^\infty e^{-pt}v(t,x)\,dt=\K (p)\int\limits_0^\infty 
e^{-up\K (p)}(4\pi u)^{-n/2}e^{-\frac{|x|^2}{4u}}\,du.
$$
By the formula 2.3.16.1 from \cite{PBM1}, the right-hand side 
coincides with the one from (4.5), so that $v(t,x)=Z(t,x)$. Now 
the non-negativity of $Z$ is a consequence of (4.23), and the 
identity (4.25) follows from (4.23), (4.24) and Fubini's theorem. 
$\qquad \blacksquare$

\bigskip
{\bf 4.3. Long time behavior.} Let us give a rigorous proof of the 
asymptotics of the mean square displacement, basic for 
applications of the distributed order calculus. We also give some 
long time estimates of the fundamental solution $Z$.

\medskip
\begin{teo}
$\mathrm{(i)}$ Let
$$
m(t)=\int\limits_{\Rn} |x|^2Z(t,x)\,dx.
$$
If $\mu (0)\ne 0$, then
\begin{equation}
m(t)\sim C\log t,\quad t\to \infty .
\end{equation}
If
\begin{equation}
\mu (\alpha )\sim a\alpha^\nu ,\quad \alpha \to 0,\quad a,\nu >0,
\end{equation}
then
\begin{equation}
m(t)\sim C(\log t)^{1+\nu },\quad t\to \infty .
\end{equation}

$\mathrm{(ii)}$ Suppose that (4.27) holds with $\nu >1$, if $n=1$, 
and with an arbitrary $\nu >0$, if $n\ge 2$. Then for $|x|\le 
\varepsilon$, $\varepsilon >0$, and $t>\varepsilon^{-1}$,
\begin{equation}
Z(t,x)\le \begin{cases}
C(\log t)^{-\frac{\nu -1}2}, & \text{if $n=1$;}\\
C|\log |x||(\log t)^{-\nu}\log (\log t), & \text{if $n=2$;}\\
C|x|^{-n+2}(\log t)^{-\nu -1}, & \text{if $n\ge 3$}.\end{cases}
\end{equation}
\end{teo}

\medskip
{\it Proof.} (i) It follows from the Plancherel identity for the 
Fourier transform that
$$
m(t)=-(2\pi )^n\left. \left( \Delta_\xi \widehat{Z}(t,\xi )\right) 
\right|_{\xi =0}.
$$
Applying the Laplace transform in $t$ we find that
$$
\widetilde{m}(p)=-(2\pi )^n\left. \left\{ \Delta_\xi 
\left( \frac{1}{p\K (p)+|\xi |^2}\right) \right\} 
\right|_{\xi =0},
$$
and after an easy calculation we get
$$
\widetilde{m}(p)=\frac{2n\cdot (2\pi )^n}{p^2\K (p)},
$$
whence
$$
m(t)=2n\cdot (2\pi )^n\int\limits_0^t\varkappa (\tau )\,d\tau
$$
where $\varkappa$ was introduced in Sect. 3.1. Now the relations 
(4.26) and (4.28) are consequences of Karamata's Tauberian theorem 
\cite{Fe}.

(ii) As before, we proceed from the integral representation (4.14) 
where the contour $S_{\gamma ,\omega }$ consists of a finite part 
$T_{\gamma ,\omega }$ and the rays $\Gamma^\pm_{\gamma ,\omega }$. 
Let $n=1$. Then (4.14) takes the form
\begin{equation}
Z(t,x)=\frac{1}2\int\limits_{S_{\gamma ,\omega }}e^{pt}\frac{\K 
(p)}{\sqrt{p\K (p)}}e^{-|x|\sqrt{p\K (p)}}\,dp.
\end{equation}

As $p\to 0$, $\sqrt{p\K (p)}\sim C(\log p^{-1})^{-\frac{1+\nu }2}$,
$\dfrac{\K (p)}{\sqrt{p\K (p)}}\sim Cp^{-1}(\log p^{-1})^{-\frac{1+\nu 
}2}$, where $\frac{1+\nu }2>1$. These asymptotic relations make 
it possible to pass to the limit in (4.30), as $\gamma \to 0$, 
substantiating simultaneously the convergence to 0 of the integral 
over $T_{\gamma ,\omega }$ and the existence of the integrals over 
the rays starting at the origin.

Thus,
\begin{equation}
Z(t,x)\le C\int\limits_0^\infty e^{rt\cos (\omega \pi 
)}r^{-1}|\log r|^{-\frac{1+\nu }2}e^{-a|x||\log r|^{-\frac{1+\nu 
}2}}\,dr.
\end{equation}
Let us decompose the integral in (4.31) into the sum of the 
integrals over $(0,1/2)$ and $(1/2,\infty )$. Estimating the 
latter we drop the factor containing $|x|$ and obtain easily the 
exponential decay, as $t\to \infty$. The integral over $(0,1/2)$ 
is estimated via the function
$$
M(t)=\int\limits_0^{1/2}e^{-art}r^{-1}\left( \log \frac{1}r\right)^{-\frac{1+\nu 
}2}\,dr.
$$

Integrating by parts we see that
$$
M(t)\le C\left( e^{-\frac{at}2}+t\int\limits_0^{1/2}e^{-art}\left( 
\log \frac{1}r\right)^{\frac{1-\nu }2}\,dr\right) .
$$
It is known (see (18.52) in \cite{Ri2} or (32.11) in \cite{Ri3}) 
that
$$
\int\limits_0^{1/2}e^{-art}\left( \log \frac{1}r\right)^{\frac{1-\nu 
}2}\,dr \le Ct^{-1}(\log t)^{\frac{1-\nu }2}
$$
for large values of $t$. This implies the first inequality of 
(4.29).

Let $n=2$. Then
$$
\widetilde{Z}(p,x)=\frac{1}{2\pi }\K (p)K_0(|x|\sqrt{p\K (p)},
$$
so that we have, for $|x|<\varepsilon$ and small $|p|$, that
$$
\left| \widetilde{Z}(p,x)\right| \le C|p|^{-1}|\log |p||^{-1-\nu 
}\left( \log |x|^{-1}+\log \log |p|^{-1}\right) .
$$

This estimate is sufficient (for $\nu >0$) to substantiate passing 
to the limit, as $\gamma \to 0$. The above argument gives, as the 
main part of the upper estimate of $Z(t,x)$ for a large $t$, the 
expression
\begin{multline*}
C\left( \log |x|^{-1}\right) \int\limits_0^{1/2}e^{-art}r^{-1}|
\log r|^{-1-\nu } \left( \log \log r^{-1}\right) \,dr\\
=C_1\left( \log |x|^{-1}\right) \int\limits_0^{1/2}e^{-art}
\left( \log \log r^{-1}\right) \frac{d}{dr}\left( \log r^{-1}\right)^{-\nu 
}\,dr.
\end{multline*}
Integrating by parts we reduce the investigation of the above 
integral in $r$ to that of two integrals,
$$
\int\limits_0^{1/2}e^{-art}r^{-1}\left( \log r^{-1}\right)^{-1-\nu 
}\,dr
$$
(it has been estimated above), and
$$
\int\limits_0^{1/2}e^{-art}\left( \log r^{-1}\right)^{-\nu }\log \log r^{-1}\,dr
\sim Ct^{-1}(\log t)^{-\nu }\log (\log t),\quad t\to \infty
$$
(\cite{Ri3}, 32.11). This results in the second estimate from 
(4.29). The third one is derived similarly. $\qquad \blacksquare$

\bigskip
The relations (4.26) and (4.28) for the case where $n=1$ and $\mu (\alpha )\equiv 
\text{const}$ or $\mu (\alpha )\equiv \text{const}\cdot \alpha^\nu$
were proved in \cite{CGSG}.

\section{The Cauchy Problem}

{\bf 5.1. The homogeneous equation.} Let us consider the equation 
(4.1) with the initial condition
\begin{equation}
u(0,x)=\varphi (x),\quad x\in \Rn,
\end{equation}
where $\varphi$ is a locally H\"older continuous function of the 
sub-exponential growth: for any $b>0$,
\begin{equation}
|\varphi (x)|\le C_be^{b|x|}.
\end{equation}
We will assume that the weight function $\mu$ defining the 
distributed order derivative $\DM$ satisfies the conditions of 
Theorem 4.1.

\medskip
\begin{teo}
$\mathrm{(i)}$ The function
\begin{equation}
u(t,x)=\int\limits_{\Rn} Z(t,x-\xi )\varphi (\xi )\,d\xi
\end{equation}
is a classical solution of the Cauchy problem (4.1)-(5.1), that is 
the function (5.3) is twice continuously differentiable in $x$ for 
each $t>0$, for each $x\in \Rn$ it is continuous in $t>0$, the 
function
$$
t\mapsto \int\limits_0^tk(t-\tau )u(\tau ,x)\,d\tau ,\quad t>0,
$$
is continuously differentiable, the equation (4.1) is satisfied, 
and
\begin{equation}
u(t,x)\longrightarrow \varphi (x),\quad \text{as }t\to 0,
\end{equation}
for all $x\in \Rn$.

$\mathrm{(ii)}$ On each finite time interval $(0,T]$, the solution 
$u(t,x)$ satisfies the inequality
\begin{equation}
|u(t,x)|\le Ce^{d|x|},\quad x\in \Rn,
\end{equation}
with some constants $C,d>0$. If $\varphi$ is bounded, then
\begin{equation}
|u(t,x)|\le C,\quad x\in \Rn,\quad 0<t\le T.
\end{equation}

$\mathrm{(iii)}$ For each $x\in \Rn$, there exists such an 
$\varepsilon >0$ that
\begin{equation}
\left| \DM u(t,x)\right| \le C_xt^{-1+\varepsilon },\quad 0<t\le T.
\end{equation}
\end{teo}

\medskip
{\it Proof}. Using (4.25) we can write
\begin{equation}
u(t,x)=\int\limits_{\Rn} Z(t,x-\xi )[\varphi (\xi )-\varphi 
(x)]\,d\xi +\varphi (x).
\end{equation}
Let us fix $x$ and prove (5.4), that is prove that the integral in 
(5.8) (denoted by $u_0(t,x)$) tends to 0.

Let $n=1$. Then
\begin{equation}
u_0(t,x)=\frac{1}{4\pi i}\int\limits_{\gamma -i\infty}^{\gamma 
+i\infty}e^{pt}\frac{\K (p)}{\sqrt{p\K (p)}}H(p,x)\,dp
\end{equation}
where $\gamma >0$,
$$
H(p,x)=\int\limits_{-\infty}^\infty e^{-|x-\xi |\sqrt{p\K (p)}}[\varphi (\xi )-\varphi 
(x)]\,d\xi
$$
(the change of the order of integration leading to (5.9) will be 
justified when we prove the decay of $H(p,x)$, as $p\to \gamma \pm 
i\infty$; see below).

By our assumption,
$$
|\varphi (x)-\varphi (\xi )|\le C_x|x-\xi |^\lambda ,\quad \lambda 
>0,\ |x-\xi |\le 1.
$$

Let $\rho =\sqrt{p\K (p)}$, $p=\gamma +i\tau$. As $|\tau |\to 
\infty$, $\rho \sim C\left( \dfrac{|\tau |}{\log |\tau 
|}\right)^{1/2}$. We have
\begin{multline*}
|H(p,x)|\le C\int\limits_{|x-\xi |\le 1}e^{-\rho |x-\xi |}|x-\xi 
|^\lambda \,d\xi +C\int\limits_{|x-\xi |>1}e^{-\rho |x-\xi |+b|\xi |} 
\,d\xi +|\varphi (x)|\int\limits_{|x-\xi |>1}e^{-\rho |x-\xi 
|}\,d\xi \\
\le C\rho^{-1-\lambda }+Ce^{b|x|}\int\limits_{|z|>1}e^{(b-\rho 
)|z|}\,dz+2|\varphi (x)|\int\limits_1^\infty e^{-\rho z}\,dz
\le C\rho^{-1-\lambda },
\end{multline*}
if $b$ is taken such that $\rho >b$. Therefore the absolute value 
of the integrand in (5.9) does not exceed
$$
Ce^{\gamma t}|\tau |^{-1-\frac{\lambda }2}(\log |\tau |)^{\lambda 
/2},
$$
so that the integral in (5.9) exists and possesses a limit, as 
$t\to 0$, equal to
\begin{equation}
u_0(0,x)=\frac{1}{4\pi i}\int\limits_{\gamma -i\infty}^{\gamma 
+i\infty}\frac{\K (p)}{\sqrt{p\K (p)}}H(p,x)\,dp
\end{equation}

The integrand in (5.10) is analytic in $p$ on the half-plane $\R 
p\ge \gamma$. Let us consider (within that half-plane) a contour 
consisting of an interval $\{ p:\ \R p=\gamma ,|p|\le R\}$ and the 
arc $\{ p:\ \R p>\gamma ,|p|=R\}$, $R>\gamma$. The absolute value of 
the integral over the arc (with the same integrand as in (5.10)) 
does not exceed $CR^{-\lambda /2}(\log R)^{\lambda /2}\to 0$, as 
$R\to \infty$. This means that $u_0(0,x)=0$, and we have proved 
(5.4) for $n=1$. The scheme of proof is completely similar for 
$n>1$ too; one has only to use the asymptotics of the McDonald 
function.

If we perform the above estimates, not ignoring the dependence on 
$x$ but, on the contrary, taking it into account, then we obtain 
the estimates (5.5) and (5.6).

Due to the estimates of $Z$ given in Theorem 4.1, we may 
differentiate once in the spatial variables in (5.3) under the 
sign of integral. Using also the identity (4.25) we get, for a 
fixed $x^0$, the formula
\begin{equation}
\frac{\partial u(t,x^0)}{\partial x_k}=\int\limits_{\Rn} \frac{\partial 
Z(t,x^0-\xi )}{\partial x_k}[\varphi (\xi )-\varphi 
(x^0)]\,d\xi .
\end{equation}

Let us decompose the domain of integration in (5.11) into the 
union of 
$$
\Omega_1=\left\{ \xi \in \Rn:\ |x^0-\xi |\ge 1\right\}
$$
and $\Omega_2=\Rn \setminus \Omega_1$. Correspondingly, the 
integral becomes a sum of two functions, $w_1(t,x)+w_2(t,x)$. If 
$x$ is in a small neighbourhood of $x^0$, while $\xi \in 
\Omega_1$, then $|x-\xi |$ is separated from zero. Therefore
\begin{equation}
\frac{\partial w_1(t,x^0)}{\partial x_k}=\int\limits_{\Omega_1} \frac{\partial^2 
Z(t,x^0-\xi )}{\partial x_k^2}[\varphi (\xi )-\varphi 
(x^0)]\,d\xi .
\end{equation}

Let $d$ be a small positive number, $\tilde{d}=(0,\ldots ,d,\ldots 
)$, with $d$ being at the $k$-th place. Then
\begin{multline}
\frac{1}d\left[ w_2(t,x^0+\tilde{d})-w_2(t,x^0)\right] -\int\limits_{\Omega_2}
\frac{\partial^2 
Z(t,x^0-\xi )}{\partial x_k^2}[\varphi (\xi )-\varphi (x^0)]\,d\xi 
\\
=\frac{1}d\int\limits_{|x^0-\xi |\le 2d} \frac{\partial 
Z(t,x^0+\tilde{d}-\xi )}{\partial x_k}[\varphi (\xi )-\varphi (x^0)]\,d\xi 
-\frac{1}d\int\limits_{|x^0-\xi |\le 2d} \frac{\partial 
Z(t,x^0-\xi )}{\partial x_k}[\varphi (\xi )-\varphi (x^0)]\,d\xi \\
-\int\limits_{|x^0-\xi |\le 2d} \frac{\partial^2 
Z(t,x^0-\xi )}{\partial x_k^2}[\varphi (\xi )-\varphi (x^0)]\,d\xi 
\\
+\int\limits_{2d\le |x^0-\xi |\le 1} \left\{ \frac{1}d\left[ \frac{\partial 
Z(t,x^0+\tilde{d}-\xi )}{\partial x_k}-\frac{\partial 
Z(t,x^0-\xi )}{\partial x_k}\right]-\frac{\partial^2 
Z(t,x^0-\xi )}{\partial x_k^2}\right\}[\varphi (\xi )-\varphi (x^0)]\,d\xi .
\end{multline}
The integrals converge due to the local H\"older continuity of 
$\varphi$.

We have (if $n\ge 2$)
\begin{multline*}
\frac{1}d\int\limits_{|x^0-\xi |\le 2d} \frac{\partial 
Z(t,x^0+\tilde{d}-\xi )}{\partial x_k}[\varphi (\xi )-\varphi (x^0)]\,d\xi 
\le Ct^{-1}d^{-1}\int\limits_{|x^0-\xi |\le 2d}|x^0+\tilde{d}-\xi |^{-n+1}
|\xi -x^0|^\lambda\,d\xi \\
=Ct^{-1}d^{-1}\int\limits_{|\eta |\le 2d}|\eta +\tilde{d}|^{-n+1}
|\eta |^\lambda\,d\eta \le Ct^{-1}d^\lambda \to 0,\quad d\to 0
\end{multline*}
(the change of variables $\eta =d\zeta$ was made in the last integral). In a 
similar way we obtain estimates of other integrals over the set 
$\{ |x^0-\xi |\le 2d\}$.

In the integral over its complement, we use the Taylor formula:
$$
\frac{1}d\left[ \frac{\partial 
Z(t,x^0+\tilde{d}-\xi )}{\partial x_k}-\frac{\partial 
Z(t,x^0-\xi )}{\partial x_k}\right]-\frac{\partial^2 
Z(t,x^0-\xi )}{\partial x_k^2}=\frac{d}2\frac{\partial^3 
Z(t,x^0+\theta \tilde{d}-\xi )}{\partial x_k^3},\quad 0<\theta <1.
$$
If $|x^0-\xi |\ge 2d$, then
$$
|x^0+\theta \tilde{d}-\xi |\ge |\xi -x^0|-d\ge \frac{1}2|\xi 
-x^0|.
$$
Using the inequality for the third derivative of $Z$ from Theorem 
4.1 we find that the last integral in (5.13) does not exceed
$$
Cdt^{-1}\int\limits_{2d\le |x^0-\xi |\le 1}|\xi -x^0|^{-n-1+\lambda 
}\,d\xi \le Ct^{-1}d^\lambda \to 0,
$$
as $d\to 0$.

It follows from (5.12), (5.13) and the above estimates that
\begin{equation}
\frac{\partial^2 u(t,x^0)}{\partial x_k^2}=\int\limits_{\Rn} \frac{\partial^2 
Z(t,x^0-\xi )}{\partial x_k}[\varphi (\xi )-\varphi 
(x^0)]\,d\xi .
\end{equation}
If $n=1$, then the formula (5.14) is obtained by a straightforward 
differentiation under the sign of integral.

Let us consider the distributed order derivative $\DM u$. First of 
all we check the identity
\begin{equation}
\DM Z(t,x)=\Delta Z(t,x),\quad t>0,\ x\ne 0.
\end{equation}
A direct calculation based on identities for the derivatives of 
the McDonald function \cite{BE} shows that $\Delta 
\widetilde{Z}(p,x)=p\K (p)\widetilde{Z}(p,x)$. On the other hand, 
if $x\ne 0$, then $Z(t,x)\to 0$, as $t\to 0$. This fact follows 
from the integral representation of $Z$ in a manner similar to the 
above proof of (5.4). Therefore the Laplace transform of $\DM 
Z(t,x)$, $x\ne 0$, equals $p\K (p)\widetilde{Z}(p,x)$, which 
implies (5.15).

Now, having the estimates of the derivatives of $Z$ in spatial 
variables given in Theorem 4.1, from (5.15) we get estimates for 
$\DM Z$ sufficient to justify the distributed differentiation in 
(5.8). Thus we come to the formula
\begin{equation}
\left( \DM u\right)(t,x^0) =\int\limits_{\Rn} \left( \DM 
Z\right) (t,x^0-\xi )[\varphi (\xi )-\varphi 
(x^0)]\,d\xi .
\end{equation}
Together with (5.14) and (5.15), this proves that $u(t,x)$ is a 
solution of the equation (4.1).

In order to prove (5.7), we use the inequalities (4.13), 
(4.13$'$), and the assumption (5.2) with $b<a$. Substituting into 
(5.16) we get, for a fixed $x^0$, that
\begin{multline*}
\left| \left( \DM u\right) (t,x^0)\right| \le 
Ct^{-2}\int\limits_{|x^0-\xi |<t^{1/2}}|x^0-\xi |^{-n+2+\lambda 
}\,d\xi+C\int\limits_{t^{1/2}\le |x^0-\xi |\le 1}|x^0-\xi |^{-n-2+\lambda 
}e^{-a|x^0-\xi |}\,d\xi \\
+C\int\limits_{|x^0-\xi |>1}|x^0-\xi |^{-n-2}
e^{-a|x^0-\xi |}\left( e^{b|\xi |}+e^{b|x^0|}\right) \,d\xi \\
\le Ct^{-2}\int\limits_0^{t^{1/2}}r^{1+\lambda 
}\,dr+C\int\limits_{t^{1/2}}^1r^{-3+\lambda }e^{-ar}\,dr+C
\le Ct^{-1+\frac{\lambda}2}
\end{multline*}
for small values of $t$, as desired. $\qquad\blacksquare$

\bigskip
{\bf 5.2. The inhomogeneous equation.} Let us consider the Cauchy 
problem
\begin{equation}
\left( \DM_t u\right) (t,x)-\Delta u(t,x)=f(t,x);\quad x\in \Rn 
,t>0,
\end{equation}
\begin{equation}
u(0,x)=0.
\end{equation}
We assume that the function $f$ is continuous in $t$, bounded and 
locally H\"older continuous in $x$, uniformly with respect to $t$. 
Our task in this section is to obtain a solution of (5.17)-(5.18) 
in the form of a ``heat potential''
\begin{equation}
u(t,x)=\int\limits_0^td\tau \int\limits_{\Rn} E(t-\tau ,x-y)f(\tau 
,y)\,dy.
\end{equation}

In contrast to the classical theory of parabolic equations 
\cite{Fri}, the kernel $E$ in (5.19) does not coincide with the 
fundamental solution $Z$ -- just as this happens for fractional 
diffusion equations \cite{EK,EIK}. However the behavior of the 
function $E$ is very similar to that of $Z$. Applying formally the 
Laplace transform in $t$ and the Fourier transform in $x$ we find 
that
$$
\widehat{\tilde{E}}(p,\xi )=\frac{1}{p\K (p)+|\xi |^2}
$$
whence
\begin{equation}
\tilde{E} (p,x)=(2\pi )^{-\frac{n}2}|x|^{1-\frac{n}2}
(p\K (p))^{\frac{1}2(\frac{n}2-1)}K_{\frac{n}2-1}(|x|\sqrt{p\K 
(p)}),
\end{equation}
which differs from (4.5) only by the absense of the factor $\K 
(p)$ with a logarithmic behavior at infinity. Therefore the 
function $E(t,x)$ obtained from (5.20) via contour integration, 
satisfies the same estimates (see Theorem 4.1) as the function 
$Z$, except the estimates for large values of $t$.

The function $E(t,x)$ is non-negative. Indeed, the function 
$p\mapsto p^{\nu /2}K_\nu (a\sqrt{p})$, $a>0$, is the Laplace 
transform of the function
$$
t\mapsto \frac{a^\nu}{(2t)^{\nu +1}}e^{-\frac{a^2}{4t}}
$$
(see \cite{DP}). This means that the above function in $p$ is 
completely monotone. Since the function $p\mapsto p\K (p)$ is 
positive and has a completely monotone derivative, we find that 
$\tilde{E}(p,x)$ is completely monotone in $p$, so that $E(t,x)\ge 
0$, $x\ne 0$.

The counterparts of the estimates (4.29) (proved just as in 
Theorem 4.3) are as follows. If (4.27) holds with $\nu \ge 0$, 
then for $|x|\le \varepsilon$, $\varepsilon >0$, and 
$t>\varepsilon^{-1}$
\begin{equation}
E(t,x)\le \begin{cases}
Ct^{-1}(\log t)^{\frac{1+\nu}2}, & \text{if $n=1$;}\\
Ct^{-1}\log \log t\log |x|^{-1}, & \text{if $n=2$;}\\
Ct^{-1}|x|^{-n+2}, & \text{if $n\ge 3$.}\end{cases}
\end{equation}
The function $E$ has (in $x$) an exponential decay at infinity.

In fact, for the analysis of the potential (5.19) we need estimates 
of $E$ and its derivatives, uniform in $t\in (0,T]$.

\medskip
\begin{prop}
Let $\mu$ satisfy the conditions of Theorem 4.1. Then, uniformly 
in $t\in (0,T]$, 
\begin{equation}
\left| D_x^jE(t,x)\right| \le C|x|^{-j-n}|1+|\log |x|||^\beta 
e^{-a|x|},\quad x\ne 0,\ j\ge 0,
\end{equation}
\begin{equation}
\left| \DM_t E(t,x)\right| \le C|x|^{-n-2}|1+|\log |x|||^\beta 
e^{-a|x|},\quad x\ne 0,
\end{equation}
whre $C,a,\beta$ are positive constants.
\end{prop}

\medskip
{\it Proof}. Let, for example, $n\ge 3$ (other cases are 
considered in a similar way). As usual, we write the Laplace 
inversion formula and deform the contour of integration to 
$S_{\gamma ,\omega}$. The integral over $T_{\gamma ,\omega}$ gives 
an exponentially decaying contribution without local 
singularities. In the integrals over the rays $\Gamma^\pm_{\gamma 
,\omega}$ we use the upper bound
$$
\left| K_{\frac{n}2-1}(z)\right| \le 
C|z|^{-\frac{n}2+1}e^{-a|z|},\quad z\ne 0,
$$
($a>0$) obtained from the asymptotics of the McDonald function 
near the origin and infinity. As in the proof of Theorem 4.1, we 
perform the change of variable $z=\left( \dfrac{r}{\log 
r}\right)^{1/2}$ and use the inequality (4.18) for the dependence 
of $r$ on $z$.

As a result, for the integrals over $\Gamma^\pm_{\gamma 
,\omega}$ we obtain the upper bound
$$
C|x|^{-n+2}\int\limits_{\gamma_1}^\infty z(\log z)^\beta 
e^{-a|x|z}\,dz\le C|x|^{-n}(|\log |x||+1)^\beta e^{-a'|x|}
$$
with some positive constants, and we come to the estimate (5.22), 
$j=0$. The estimates of the derivatives in spatial variables are 
proved similarly.

The proof of (5.23) is completely analogous to that of the 
inequality (4.13) for $\DM Z$. $\qquad \blacksquare$

\medskip
As we have noticed,
$$
\tilde{E}(p,x)=\frac{1}{\K (p)}\tilde{Z}(p,x),
$$
and since
$$
\int\limits_{\Rn}\tilde{Z}(p,x)\,dx=\frac{1}p,
$$
we have
$$
\int\limits_{\Rn}\tilde{E}(p,x)\,dx=\frac{1}{p\K (p)},
$$
so that we come to an interesting identity
\begin{equation}
\int\limits_{\Rn}E(t,x)\,dx=\varkappa (t).
\end{equation}

The existence of the integral in (5.24) follows from the above 
estimates or from the fact that $\varkappa \in L_1^{\text{loc}}$ 
(see (3.2)) and Fubini's theorem.

\medskip
\begin{teo}
Under the above assumptions regarding $f$, and the assumptions of 
Theorem 4.1 regarding $\mu$, the function (5.19) is a classical 
solution of the problem (5.17)-(5.18), bounded near the origin in 
$t$ for each $x\in \Rn$.
\end{teo}

\medskip
{\it Proof.} The initial condition (5.18) is evidently satisfied. 
Just as for the kernel $Z$ above, we prove that $\DM E-\Delta E=0$ 
for $x\ne 0$. Next, we may differentiate once in (5.19) under the 
sign if integral. Indeed (here and below we make estimates for 
$n\ge 3$; other cases are similar),
\begin{multline*}
\int\limits_0^td\tau \int\limits_{\Rn}\left| \frac{\partial 
E(t-\tau ,x-y)}{\partial x_j}\right| \,dy
\le 
C\int\limits_0^td\tau \int\limits_{|x|>\sqrt{\tau}}|x|^{-n-1}(1+|\log 
|x||)^\beta e^{-a|x|}\,dx\\
+C\int\limits_0^t\tau^{-1}d\tau \int\limits_{|x|\le 
\sqrt{\tau}}|x|^{-n+1}\,dx \\
\le C\int\limits_0^t\tau^{-1/2}d\tau 
\int\limits_{|y|>1}|y|^{-n-1}(1+\log |y|+\frac{1}2\log 
\tau^{-1})\,dy +C\int\limits_0^t\tau^{-1/2}d\tau 
\int\limits_{|y|\le 1}|y|^{-n+1}\,dy<\infty.
\end{multline*}

In order to calculate the second order derivatives, note that the 
function
$$
u_h(t,x)=\int\limits_0^{t-h}d\tau \int\limits_{\Rn} E(t-\tau ,x-y)f(\tau 
,y)\,dy,\quad t>h>0,
$$
may be differentiated twice, and that
$$
\int\limits_{\Rn}\frac{\partial^2}{\partial x_i^2}E(t-\tau 
,x-y)\,dy=0,
$$
whence
\begin{equation}
\frac{\partial^2u_h(t,x)}{\partial x_i^2}=\int\limits_0^{t-h}d\tau 
\int\limits_{\Rn}\frac{\partial^2}{\partial x_i^2} E(t-\tau ,x-y)[f(\tau 
,y)-f(\tau ,x)]\,dy.
\end{equation}

Using the local H\"older continuity and boundedness of $f$, we 
perform estimates as above and prove the possibility to pass to 
the limit in (5.25), as $h\to 0$, so that
\begin{equation}
\Delta u(t,x)=\int\limits_0^td\tau \int\limits_{\Rn}\Delta E(t-\tau ,x-y)[f(\tau 
,y)-f(\tau ,x)]\,dy.
\end{equation}

To calculate $\DM u$, we use (5.24) and write
$$
u(t,x)=\int\limits_0^td\tau \int\limits_{\Rn}E(t-\tau ,x-y)[f(\tau 
,y)-f(\tau ,x)]\,dy+\int\limits_0^t\varkappa (t-\tau )f(\tau 
,x)\,d\tau \overset{\text{def}}{=} u_1(t,x)+u_2(t,x).
$$
Recall that $u_2(t,x)=\left( \IM f\right) (t,x)$, so that $\DM 
u_2=f$ (see Sect. 3).

Let us consider $u_1$. First we estimate $\dfrac{\partial 
E}{\partial t}$. As before, we use the contour integral 
representation of $E$ and note that the differentiation in $t$ 
leads to an additional factor $p$ in the integrals. This results 
in the estimates
\begin{equation}
\left| \frac{\partial E(t,x)}{\partial t}\right| \le 
Ct^{-2}|x|^{-n+2},\quad |x|\le \varepsilon;
\end{equation}
\begin{equation}
\left| \frac{\partial E(t,x)}{\partial t}\right| \le 
C|x|^{-n-2}(|\log |x||+1)^\beta e^{-a|x|},\quad |x|\ne 0.
\end{equation}

As the first step of computing $\DM u_1$, we compute $\dfrac{\partial 
u_1}{\partial t}$. Note that
\begin{equation}
\int\limits_{\Rn}E(t-\tau ,x-y)[f(\tau ,y)-f(\tau 
,x)]\,dy=\frac{1}{(2\pi )^{n/2+1}i}\int\limits_{\gamma 
-i\infty}^{\gamma +i\infty}e^{p(t-\tau )}(p\K 
(p))^{\frac{1}2(\frac{n}2-1)}L_n(p,x,\tau )\,dp
\end{equation}
where
$$
L_n(p,x,\tau )= \int\limits_{\Rn}|x-y|^{1-\frac{n}2}[f(\tau ,y)-f(\tau 
,x)]K_{\frac{n}2-1}(|x-y|\sqrt{p\K (p)})\,dy.
$$

The role of the function $L_n$ is quite similar to that of the 
function $H$ introduced in the proof of Theorem 5.1 (where the 
case $n=1$ was considered in detail). Using, as it was done there, 
the local H\"older continuity and boundedness of $f$ we find that
$$
|L_n(p,x,\tau )|\le C|\sqrt{p\K (p)}|^{-\frac{n}2-\lambda -1}.
$$
As in the proof of Theorem 5.1, we deform the contour of 
integration to the right of the line in (5.29) and show that
\begin{equation}
\lim\limits_{\tau \to t}\int\limits_{\Rn}E(t-\tau ,x-y)[f(\tau ,y)-f(\tau 
,x)]\,dy=0.
\end{equation}

On the other hand, using (5.27) and (5.28) we get
\begin{multline*}
\int\limits_{\Rn}\left| \frac{\partial E(t-\tau ,x-y)}{\partial 
t}\right| |f(\tau ,y)-f(\tau ,x)|\,dy
\le C\int\limits_{|y|\ge \sqrt{t-\tau }}|y|^{-n-2+\lambda}(|\log |y||+1)^\beta 
e^{-a|y|}\,dy\\
+C(t-\tau )^{-2}\int\limits_{|y|<\sqrt{t-\tau 
}}|y|^{-n+2+\lambda}\,dy=2C\int\limits_{\sqrt{t-\tau }}^\infty r^{-3+\lambda } 
(|\log r|+1)^\beta e^{-ar}\,dr+2C(t-\tau )^{-2}\int\limits_0^{\sqrt{t-\tau 
}}r^{1+\lambda }\,dr\\
\le C(t-\tau )^{-1+\lambda /2}(|\log (t-\tau )|+1)^\beta .
\end{multline*}
Together with (5.30), this implies the equality
$$
\frac{\partial u_1}{\partial t}=\int\limits_0^td\tau 
\int\limits_{\Rn}\frac{\partial E(t-\tau ,x-y)}{\partial t}[f(\tau ,y)-f(\tau 
,x)]\,dy.
$$

Now we compute $\DM u_1$ using the formula (2.3), the fact that 
$k\in L_1^{\text{loc}}$ (following from (2.4)) and Fubini's 
theorem:
$$
\left( \DM u_1\right) (t,x)=\int\limits_0^td\tau 
\int\limits_{\Rn}\left( \DM E\right) (t-\tau ,x-y)[f(\tau ,y)-f(\tau 
,x)]\,dy.
$$
Together with (5.26), this means that $\Delta u=\DM u_1=\DM u-f$, 
as desired. $\qquad \blacksquare$

\medskip
In Theorem 5.3 we constructed a solution $u$ of the problem 
(5.17)-(5.18), such that $u=u_1+u_2$, $u_1(0,x)=u_2(0,x)=0$, $u_1$ 
is absolutely continuous in $t$, and $u_2=\IM f$. On this solution 
$u$,
$$
\IM \DM u=\IM (k*u_1')+\IM f=\varkappa *k*u_1'+u_2=u_1+u_2=u
$$
($u'$ means the derivative in $t$). Applying $\IM$ to both sides 
of the equation (5.17) we find that
\begin{equation}
u(t,x)-\int\limits_0^t\varkappa (t-s)\Delta u(s,x)\,ds=(\varkappa 
*f)(t,x).
\end{equation}

The equation (5.31) can be interpreted as an abstract Volterra 
equation
\begin{equation}
u+\varkappa *(Au)=\varphi ,
\end{equation}
if we assume that $u$ belongs to some Banach space $X$ (in the 
variable $x$), and $A$ is the operator $-\Delta$ on $X$. The 
operator $-A$ generates a contraction semigroup if, for example, 
$X=L_2(\Rn )$ or $X=C_\infty (\Rn )$ (the space of continuous 
functions decaying at infinity; see Sect. X.8 in \cite{RS}). Now 
the existence of a solution in $L_1(0,T;X)$ can be obtained from a 
general theory of equations (5.32) developed in \cite{CN}; it is 
essential that $\varkappa$ is completely monotone (conditions of 
some other papers devoted to equations (5.32) do not cover our 
situation). Of course, our ``classical'' approach gives a much 
more detailed information about solutions, while the abstract 
method is applicable to more general equations.

\bigskip
\section{Uniqueness Theorems}

{\bf 6.1. Bounded solutions.} In this section we consider a more 
general equation
\begin{equation}
\left( \DM u\right) (t,x)=Lu(t,x),\quad x\in \Rn ,\ 0<t\le T,
\end{equation}
with the zero initial condition
\begin{equation}
u(0,x)=0.
\end{equation}
Here $L$ is an elliptic second order differential operator with 
bounded continuous real-valued coefficients:
$$
Lu=\sum\limits_{i,j=1}^na_{ij}(t,x)\frac{\partial^2u}{\partial 
x_i\partial x_j}+\sum\limits_{j=1}^nb_j(t,x)\frac{\partial u}{ 
\partial x_j}+c(t,x)u,
$$
$$
\sum\limits_{i,j=1}^na_{ij}(t,x)\xi_i\xi_j>0,\quad 0\ne \xi 
=(\xi_1,\ldots ,\xi_n)\in \Rn .
$$
We assume that $\mu \in C^3[0,1]$, $\mu (1)\ne 0$.

We will consider classical solutions $u(t,x)$, such that $\left(
\DM u\right) (t,x)$ belongs, for each fixed $x$, to $L_p(0,T)$ 
with some $p>1$. As we saw in Theorem 5.1 and Theorem 5.3, the 
solutions for the case $L=\Delta$ obtained via the fundamental 
solution and heat potential possess the last property making it 
possible to represent $\DM u$ in the Marchaud form (3.9).

It is often convenient to transform the equation (6.1) setting
$$
u(t,x)=u_\lambda (t)w(t,x)
$$
where $\lambda >0$, and $u_\lambda$ is the solution of the 
equation $\DM u_\lambda =\lambda u_\lambda$ constructed in Sect. 
2.3. It is easy to check that the function $w$ satisfies the 
equation
$$
\left( A_\lambda w\right) (t,x)=(L-\lambda )w(t,x)
$$
where
\begin{equation}
\left( A_\lambda w\right) (t,x)=\frac{1}{u_\lambda (t)}\left\{ 
k(t)w(t,x)+\lim\limits_{\varepsilon \to 
0}\int\limits_0^{t-\varepsilon }u_\lambda (\tau )k'(t-\tau 
)[w(\tau ,x)-w(t,x)]\,d\tau \right\} .
\end{equation}
The operator (6.3) is very similar in its properties to the 
distributed order derivative $\DM$.

\medskip
\begin{teo}
If $u(t,x)$ is a bounded classical solution of the problem 
(6.1)-(6.2), such that for each $x\in \Rn$, $\DM u\in L_p(0,T)$ 
for some $p>1$, then $u(t,x)\equiv 0$.
\end{teo}

\medskip
{\it Proof}. Let $M=\sup |u(t,x)|$. Consider the function
$$
F_R(t,x)=\frac{M}{R^2}\left( |x|^2+\sigma \int\limits_0^t\varkappa 
(s)\,ds+1\right) ,
$$
with $R,\sigma >0$. It follows from (3.2) that $\int\limits_0^t\varkappa 
(s)\,ds\to 0$, as $t\to 0$. As we have seen (Sect. 3.2), 
$\DM_t\left( \int\limits_0^t\varkappa (s)\,ds\right) =1$, so that
$$
\left( \DM F_R\right) (t,x)=\frac{\sigma M}{R^2}.
$$

Let $c_0=\sup |c(t,x)|$, $d>0$, $\lambda =c_0+d$. Since 
$u_\lambda$ is non-decreasing (Theorem 2.3), and $k'(s)\le 0$, we 
have
$$
\left( A_\lambda F_R\right) (t,x)\ge \frac{\left( \DM F_R\right) 
(t,x)}{u_\lambda (T)}=\frac{\sigma M}{R^2u_\lambda (T)}.
$$
On the other hand,
$$
\left( LF_R\right) (t,x)\le \frac{2M}{R^2}\left( 
\sum\limits_{i=1}^na_{ii}(x)+\sum\limits_{j=1}^nb_j(x)x_j\right) 
\le \frac{2M}{R^2}\left( C_1+C_2|x|\right) ,\quad C_1,C_2>0,
$$
so that taking $d>0$ we get
$$
\left( \left( A_\lambda -(L-\lambda)\right) F_R\right) (t,x)\ge 
\frac{M}{R^2}\left( \frac{\sigma}{u_\lambda 
(T)}-2C_1-2C_2|x|+d|x|^2+d\right) \ge 0
$$
for all $x\in \Rn$, $t\in (0,T)$, if $\sigma$ is taken 
sufficiently big.

Denote $v(t,x)=u(t,x)-F_R(t,x)$. By the above inequalities,
\begin{equation}
\left( A_\lambda v\right) (t,x)-(L-\lambda)v(t,x)\le 0.
\end{equation}
If $|x|=R$, then $v(t,x)=u(t,x)-M-R^{-2}\left( \sigma \int\limits_0^t\varkappa 
(s)\,ds+1\right)<0$. Next, $v(0,x)=-F_R(0,x)<0$ for all $x$. This 
means that $v(t,x)\le 0$ for $|x|<R$, $t\in [0,T]$. Indeed, 
otherwise the function $v$ would possess a point of the global 
maximum $(t^0,x^0)$ on the set $\{ (t,x)|\ 0<t\le T,|x|<R\}$, such 
that $v(t^0,x^0)>0$. Then
$$
(L-\lambda )v(t^0,x^0)\le 0
$$
(see the proof of the maximum principle for a second order 
parabolic differential equation \cite{Fri,Lan}), so that 
$\left( A_\lambda v\right) (t^0,x^0)\le 0$, due to (6.4). However 
it follows from (6.3) that
$\left( A_\lambda v\right) (t^0,x^0)>0$, and we have come to a 
contradiction.

Thus, we have proved that
$$
u(t,x)\le \frac{M}{R^2}\left( |x|^2+\sigma \int\limits_0^t\varkappa 
(s)\,ds+1\right) ,\quad |x|\le T.
$$
Since $R$ is arbitrary, we find that $u(t,x)\le 0$ for all $t\in 
[0,T]$, $x\in \Rn$. Considering $-u(t,x)$ instead of $u(t,x)$ we 
prove that $u(t,x)\equiv 0$. $\qquad \blacksquare$

\medskip
The above proof was based on standard ``maximum principle'' 
arguments. In fact, it is easy to prove, for the equation (6.1), 
an analog of the maximum principle itself. Namely, let 
$c(t,x)-\lambda \le 0$ for $t\in [0,T]$, $x\in G$, where $G\subset 
\Rn$ is a bounded domain. Suppose that
$$
(L-\lambda )u(t,x)-\left( A_\lambda u\right) (t,x)\ge 0,\quad 
(t,x)\in [0,T]\times \overline{G}.
$$
Then, if $\sup\limits_{[0,T]\times \overline{G}}u>0$, then
$$
\sup\limits_{[0,T]\times \overline{G}}u=\sup\limits_{[0,T]\times 
\partial G}u.
$$
The proof is similar to the classical one \cite{Lan}.

\bigskip
{\bf 6.2. Solutions of subexponential growth.} In this section we 
will prove a more exact uniqueness theorem for the case where 
$n=1$, $L=\dfrac{\partial^2}{\partial x^2}$.

\medskip
\begin{teo}
Suppose that $u(t,x)$ is a classical solution of the problem 
(6.1)-(6.2) with $n=1$, $L=\dfrac{\partial^2}{\partial x^2}$, such 
that for any $a>0$, 
$$
|u(t,x)|\le C_ae^{a|x|},\quad 0<t\le T,\ x\in \mathbb R^1,
$$
and $\DM_t u\in L_p(0,T)$, $p>1$, in $t$, for any fixed $x$. Then
$u(t,x)\equiv 0$.
\end{teo}

\medskip
{\it Proof}. This time we choose the comparison function as
\begin{equation}
F_R^{(1)}(t,x)=Me^{aR}[Z(t,x-R)+Z(t,x+R)],\quad |x|\le R,
\end{equation}
where $Z$ is the above fundamental solution of the Cauchy problem 
(Sect. 4), $M$ and $a$ are positive constants to be specified 
later. We will need the following auxiliary result.

\medskip
\begin{lem}
For any $T>0$, there exists such a constant $\rho_0>0$ that
$$
Z(t,0)\ge \rho_0,\quad 0<t\le T.
$$
\end{lem}

\medskip
{\it Proof}. By (4.7),
$$
\widetilde{Z}(p,0)=\frac{1}2\sqrt{p^{-1}\K (p)},
$$
so that
\begin{equation}
\widetilde{Z}(p,0)\sim \frac{\sqrt{\mu (1)}}2(p\log 
p)^{-1/2},\quad p\to \infty .
\end{equation}
Note that in Sect. 4 we used the Laplace inversion formula for 
$Z(t,x)$ only for $x\ne 0$. Here the task is just the opposite, 
and we use the inversion formula from \cite{Evg} involving the 
derivative of the Laplace image. In our case
$$
\frac{\partial \widetilde{Z}(p,0)}{\partial p}=\frac{1}{2p}\left( 
\frac{d}{dp}\sqrt{\K (p)}\cdot \sqrt{p}-\frac{\sqrt{\K 
(p)}}{2\sqrt{p}}\right) ,\quad \frac{d}{dp}\sqrt{\K 
(p)}=\frac{\K'(p)}{2\sqrt{\K (p)}},
$$
$$
\K'(p)=\int\limits_0^1(\alpha -1)p^{\alpha -2}\mu (\alpha 
)\,d\alpha =o(p^{-1}),\quad p\to \infty ,
$$
so that
$$
\left| \frac{\partial \widetilde{Z}(p,0)}{\partial p}\right| \le 
C_\varepsilon |p|^{-\frac{3}2+\varepsilon },\quad \R p\ge 1,
$$
for any $\varepsilon >0$. This is sufficient for the inversion 
formula
\begin{equation}
Z(t,0)=\frac{1}{2\pi i}\int\limits_{\gamma -i\infty }^{\gamma 
+i\infty }\widetilde{Z}(p,0)e^{pt}\,dp,\quad t\ne 0,
\end{equation}
where $\gamma \ge 1$. Using (6.6), (6.7) and an asymptotic theorem 
for the Laplace inversion (see (22.115) and (22.114) in 
\cite{Ri2}) we find the asymptotics
\begin{equation}
Z(t,0)=Ct^{-1/2}\left( \log \frac{1}t\right)^{-3/2},\quad t\to +0.
\end{equation}
For our purpose, it is sufficient to derive from (6.8) that 
$Z(t,0)\to +\infty$, as $t\to +0$.

On the other hand, it follows from the subordination identity 
(4.23) and the fact that $G(u,t)>0$ for each $t$ on the set of a 
positive measure in $u$ (see (4.24)), that $Z(t,0)>0$ for each 
$t>0$. Together with (6.8), this implies the required inequality. 
$\qquad \blacksquare$

\medskip
{\it Proof of Theorem 6.2 (continued)}. If $|x|=R$, that is $x=\pm 
R$, then by (6.5) and Lemma 6.3,
$$
F_R^{(1)}(t,x)=Me^{aR}[Z(t,0)+Z(t,\pm 2R)]\ge M\rho_0e^{aR}.
$$
We have $u(t,x)\le F_R^{(1)}(t,x)$, if $C_a\le M\rho_0$ ($a$ has 
not yet been chosen), that is if $M$ is chosen in such a way that 
$M\ge C_a\rho_0^{-1}$.

The function $w(t,x)=F_R^{(1)}(t,x)-u(t,x)$ satisfies, if $|x|\le 
R$, $t\in (0,T)$, the equation 
$\DM_tw=\dfrac{\partial^2w}{\partial x^2}$. If $|x|=R$, then 
$w(t,x)\ge 0$. In addition,
$$
\liminf\limits_{t\to 0}w(t,x)\ge 0,
$$
if $|x|<R$. It follows that $w(t,x)\ge 0$ for $t\in (0,T]$, 
$|x|\le R$.

Indeed, if $w(t,x)<0$ for some $t$ and $x$, then there exist 
$t^0\in (0,T]$, $x^0\in \mathbb R^1$, $|x^0|<R$, such that
$$
w(t^0,x^0)=\inf\limits_{\substack{|x|\le R\\ t\in (0,T]}}w(t,x)<0.
$$
If $|x|<R$, then the function $F_R^{(1)}$ is infinitely 
differentiable in $t$, with the derivative being continuous on 
$[0,T]$. Therefore we may write $\DM w$ in the Marchaud form, so 
that
\begin{equation}
k(t)w(t,x)+\lim\limits_{\varepsilon \to 
0}\int\limits_\varepsilon^t k'(\tau )[w(t-\tau ,x)-w(t,x)]\,d\tau 
=\frac{\partial^2w(t,x)}{\partial x^2}.
\end{equation}

For $(t,x)=(t^0,x^0)$, we see that the left-hand side of (6.9) is 
negative, while the right-hand side is non-negative, and we get a 
contradiction. Thus, we have proved that
\begin{equation}
u(t,x)\le Me^{aR}[Z(t,x-R)+Z(t,x+R)],\quad 0<t\le T,\ |x|\le R.
\end{equation}

Let us fix $x$ and consider the limit $R\to \infty$. For large 
values of $R$ we have
$$
Z(t,x\pm R)\le Be^{-bR},\quad t\in (0,T],
$$
where $b>0$ depends only on $T$, $B>0$ depends on $T$ and $x$, and 
not on $R$. By (6.10),
\begin{equation}
u(t,x)\le BMe^{(a-b)R}.
\end{equation}
Now we choose $a$ in such a way that $a<b$, and (see above) fix 
$M\ge C_a\rho_0^{-1}$. Obviously, $M$ does not depend on $R$. 
Passing to the limit in (6.11), as $R\to \infty$, we see that 
$u(t,x)\le 0$ for arbitrary $t$ and $x$. Similarly, taking 
$-u(t,x)$ instead of $u(t,x)$, we find that $u(t,x)\ge 0$. $\qquad 
\blacksquare$

\end{document}